\documentclass[a4paper,twoside,12pt]{article}
\usepackage{url}
\usepackage{natbib} 
\usepackage[T1]{fontenc}
\usepackage{palatino}
\usepackage{graphics}
\usepackage{graphicx}
\usepackage{booktabs,caption}
\usepackage{threeparttable}
\usepackage{adjustbox}
\usepackage[format=plain, labelsep=period, labelfont=bf, font=footnotesize,singlelinecheck=false,skip=5pt]{caption}
\usepackage{tabularx}
\usepackage{array}
\newcolumntype{H}{>{\setbox0=\hbox\bgroup}c<{\egroup}@{}}
\usepackage{amsthm}
\usepackage{lscape}
\usepackage{appendix}
\usepackage[bookmarks]{hyperref}
\usepackage{subcaption}
\usepackage{epstopdf}
\usepackage{amsmath}

\usepackage{multirow}
\usepackage{setspace}
\usepackage{amssymb}
\usepackage{textcomp}
\usepackage{amsfonts}
\usepackage{soul}
\usepackage{bbm}
\usepackage{multirow}
\usepackage{multicol}
\usepackage{verbatim}
\usepackage{mdwlist}
\usepackage{color,soul}
\usepackage{eurosym}
\usepackage{rotating}
\usepackage{lmodern}
\usepackage{color}
\usepackage{longtable}
\usepackage{etoolbox}
\usepackage{epigraph}
\usepackage{xargs}
\usepackage[left=2cm,right=2cm,bottom=2cm,top=2cm]{geometry}
\usepackage{ifpdf}
\usepackage[colorinlistoftodos,prependcaption, color=yellow,textsize=normalsize]{todonotes}
\setuptodonotes{inline,fancyline, color=yellow}

\usepackage{mathtools} 

\title{Let's roll back! The challenging task of regulating temporary contracts}

\author{Davide Fiaschi\thanks{University of Pisa, Dipartimento di Economia e Management, University of Pisa, REMARC and Centro DAGUM, Via Ridolfi 10, 56124 Pisa (Italy), Phone: +39 0502216208, Email: davide.fiaschi@unipi.it. }  \and Cristina Tealdi\thanks{Corresponding author. Edinburgh Business School, Heriot-Watt University, EH14 4AS Edinburgh (UK) and IZA Institute of Labor, Phone: +44 0131 4513803, Email: c.tealdi@hw.ac.uk.}
}

\date{\today}	

\begin{document}

\maketitle

\begin{abstract}
In this paper, we evaluate the impact of a reform introduced in Italy in 2018 (Decreto Dignità), which increased the rigidity of employment protection legislation (EPL) of temporary contracts, rolling back previous policies, to reduce job instability. We use longitudinal labour force data from 2016 to 2019 and adopt a time-series technique within a Rubin Casual Model (RCM) framework to estimate the causal effect of the reform. We find that the reform was successful in reducing persistence into temporary employment and increasing the flow from temporary to permanent employment, in particular among women and young workers in the North of Italy, with significant effects on the stocks of permanent employment (+), temporary employment (-) and unemployment (-). However, this positive outcome came at the cost of higher persistence into inactivity, lower outflows from unemployment to temporary employment and higher outflows from unemployment to inactivity among males and low-educated workers. 
\end{abstract}



\clearpage

\section{Introduction}

In several Mediterranean countries the  deregulation of temporary contracts during the 80s and 90s led to a significant increase in their utilization and to the surge of dual labour markets \citep{saint1996dual}. More recently, however, several governments (e.g., the Spanish government in 2022) have tried to roll back some of the reforms with the aim of reducing labor market segmentation \citep{bentolila2019dual}. Specifically, more restrictions have been introduced for the utilization of temporary contracts with the objective to push firms towards hiring workers under permanent contracts.
However, the literature shows that these attempts to bring the labour markets back to pre-reform arrangements may not necessarily lead to a reduction in labour market segmentation, and may even bring about unexpected outcomes (see e.g.,  \cite{cahuc2022employment}). 

In this paper we investigate the effect of a labour market reform  which was implemented in Italy in 2018, the \textit{Decreto Dignit\`a}, which increased the stringency of employment protection for temporary contracts. The reform was very controversial as policy makers, union representatives and political parties had contrasting opinions about its potential impact. The Italian social security institute estimated that  8000 temporary workers could lose their temporary jobs without finding new employment. These predictions were immediately dismissed by the Labour and Economics Ministers who deemed them “unscientific and disputable”.

Our aim is to gauge the success of the reform in reducing persistence in temporary employment, and increasing the transitions from temporary to permanent employment, exploring potential negative side effects on unemployment and labour force participation.
To achieve this goal, we use a methodology for causal policy evaluation proposed by \cite{menchetti2023combining}. Specifically, we estimate the causal-ARIMA (C-ARIMA) model, within the Rubin Casual Model (RCM) framework, which is suitable to address our research question as it  allows for the identification of the causal effects of the reform, given that the intervention is universal, non-temporary and applies to all individuals. This approach is based on the key assumptions that there is no anticipation effect, the policy is single and persistent, there is no temporal interference and the process is conditional stationary \citep{menchetti2023combining}. In Section \ref{sec:DecretoDignitàtiming} we document that these conditions are fulfilled. 

While the literature on the impact of the deregulation of temporary contracts is abundant \citep{Blanchard_Portugal_2001, blanchard2002perverse, cahuc2002temporary, cahuc2016explaining, berton2012workers, felgueroso2018surge, faccini2014reassessing}, studies of the impact of the increased stringency of EPL for temporary contracts are lacking. To the best of our knowledge, the only work which investigates the effects of a reform implemented to reduce the utilization of temporary contracts is  the paper of \cite{cahuc2022employment} for the case of Portugal. However, the specific reform they analyze applied only to large firms (above 150 employees) and not to the entire universe of firms, thus excluding the possibility to evaluate the impact on the labour market as a whole.

We use longitudinal quarterly labour force data for the period 2016-2019.  We report transition probabilities between five labour market states: \textit{inactivity} (NLFET), \textit{unemployment} (U), \textit{fixed-term employment} (FT), \textit{permanent employment} (PE) and \textit{self-employment} (SE), and the corresponding shares before and after the implementation of the \textit{Decreto Dignit\`a} reform. We then compare the forecasted values with the observed ones to evaluate the success of the reform. 
We find that after the reform, transitions probabilities from temporary employment to permanent employment increased of about 8.1 percentage points, while persistence in temporary employment decreased by 6.6 percentage points. That is, the goals of the reform were fully achieved. These effects were particularly strong among women (+10.5 pp, -7.4 pp), young (+9.1 pp, -7.6 pp) and workers in the North of Italy (+11 pp, -8.9 pp). We also provide evidence of a lower share of workers in unemployment, as a result of the reform. We discuss this effect to be due to a compositional change in the allocation of workers across contracts rather than a change in the transition probabilities to and from unemployment. The positive outcomes just discussed however came at a cost: for males and low-educated workers we estimate significantly lower outflow probabilities from unemployment to temporary employment and significantly higher outflows towards inactivity in the four quarters following the implementation of the reform..
 
The paper is organized as follows. Section \ref{sec:methodology} explains in details the proposed methodology, which is based on a view of the labour market in terms of flows across states. Section \ref{sec:empirics} describes the institutional background and the content and timing of the \textit{Decreto Dignit\`a}, while Section \ref{sec:dataset} presents the data.  Section \ref{sec:evaluationDecretoDignità}  discusses the results and finally, Section \ref{sec:concludingRemarks} gathers some concluding remarks.

\section{Methodology}\label{sec:methodology}

We measure  labour market dynamics focusing on observed transitions of working age individuals between labour market states, e.g. employment, unemployment, etc. We first describe the methodology to estimate the transition probabilities (section \ref{sec:transitionK}) and then we explain how we perform the evaluation of the policy reform using a time-series technique within a Rubin Casual Model (RCM) framework \citep{menchetti2023combining} (Section \ref{sec:counterfactualViaARIMA}). 

\subsection{The estimate of transitions with K states}\label{sec:transitionK}

In a  setting with $K$ labour market states, we assume that the labour market dynamics between periods $t-1$ and $t$ can be expressed as follows:
\begin{equation}
\boldsymbol{\pi}_t = \boldsymbol{\pi}_{t-1}\mathbf{M}_t,
\label{eq:dynamicsStocks}
\end{equation} 
where $\boldsymbol{\pi}$ is a $ 1\times K $ vector collecting the shares of individuals in the working age population in different $K$ states, and $\mathbf{M}$ is a  $K \times K$ matrix, whose elements are the transition probabilities between different states, with the constraint that:
\begin{equation}
\boldsymbol{\pi}_t\mathbf{1}^T  = 1 \; \forall t,
\label{eq:constantWorkingAgePop}
\end{equation}
where $\mathbf{1}$ is a  $1 \times K $ vector of ones; Equation (\ref{eq:constantWorkingAgePop}) simply states that the shares of working age individuals in the $K$ labour market states sum to one.

Each element of the  matrix of  transition probabilities $\mathbf{M_t}$ is assumed to satisfy the following conditions:
\begin{eqnarray} 
	\begin{cases}
		m\left(i,j\right)_t  \geq 0  \; \forall i,j \in \{1,..., K\}; \; \text{and} \\
		\sum_{j=1}^{K} m\left(i,j\right)_t  =  1 \; \forall i\in \{1,..., K\},
	\end{cases}
\label{eq:hypothesisOnQ}
\end{eqnarray}
which amount to assume that the process governing the labour market dynamics is \textit{conservative} \citep[p. 180]{cox1977theory}, i.e., there are no entries and exits from/to the working age population and, hence, the working age population is constant between $t-1$ and $t$. 

The estimate of $m\left(i,j\right)$ needs to track the status of individuals over time, i.e., we should dispose of longitudinal labour force data in which the frequency of observation (yearly, quarterly) decides the time scale of the analysis. \citet[p. 92]{anderson1957statistical} show that $\hat{m}\left(i,j\right)_{t}$, defined as:
\begin{equation}
\hat{m}\left(i,j\right)_{t} = \dfrac{M\left(i,j\right)_{t}}{M\left(i\right)_{t-1}},
\end{equation}
where $M\left(i,j\right)_{t}$ is the number of transitions between states $i$ and $j$ in the period $[t,t-1]$ and $M\left(i\right)_{t-1}$ is the total number of observations in state $i$ at time $t-1$, is the maximum likelihood estimate of $m\left(i,j\right)_{t}$ and the asymptotic distribution of such estimate is normal.
  
\subsection{The causal-ARIMA approach \label{sec:counterfactualViaARIMA}}

To estimate the effect of the \textit{Decreto Dignit\`a} reform, we use the causal-ARIMA (C-ARIMA) methodology proposed by \cite{menchetti2023combining}, which is  suitable to address our research question as it  allows for the identification of the causal effect of the reform, when the intervention is \textit{erga omnes}.\footnote{This is the main difference with respect to the more general causal effect defined in \cite{bojinov2019time}. See Footnote 6 in \cite{menchetti2023combining}.}
 In particular, under the Rubin Casual Model (RCM), the casual effect of an intervention is typically measured by the difference between the \textit{potential outcome} of the ``treated'' group and that of the ``control'' group, where only one potential outcome is observed while the other is missing and becomes counterfactual once the treatment is assigned. Within this framework, Difference-in-Differences (DiD) and synthetic controls methods are commonly used for policy evaluation. However, both methodologies require the presence of a control group, which is not subject to the treatment. Instead, the C-ARIMA approach, exploiting a novel time-series technique within a RCM framework, allows for the estimation of the causal effect of an intervention when no control group is available, given a number of assumptions being satisfied. First, all units need to be subject at the same time to a \textit{single and persistent intervention}. This implies the irreversibility of the treatment assumption  \citep[p. 5]{callaway2021difference}, i.e., once a unit becomes treated, that unit will remain treated in all future periods. Second, the \textit{temporal no-interference} assumption must hold, which is known as  the temporal stable unit treatment value assumption (TSUTVA) \citep{bojinov2019time}. This is the time series equivalent of the cross-sectional SUTVA and implies that the potential outcome only depends upon each unit individual’s treatment path. Third,  the \textit{no anticipatory effects} assumption must be fulfilled, i.e. the statistical units need to have no expectations about future interventions \citep[p. 5]{menchetti2023combining}.
 Moreover, in the presence of covariates,  the assumption of \textit{covariates treatment independence} must be satisfied, i.e., the covariates should not be affected by the intervention \citep[p. 6]{menchetti2023combining}.
Conditional on all  previously mentioned assumptions being satisfied, the potential outcome of  ``non-treated'' individuals can be estimated by forecasting a time-series model (e.g., the ARIMA model). This forecast provides a correct estimate under the final assumption of \textit{conditional stationary} of the data generation process of the potential outcome of the non-treated group.  In other words, the model fitted prior to the intervention approximates the distribution of the potential outcome of the "non-treated" group after the intervention. 

In particular, the $f$-quarter ahead forecast of transition probability in period $t$ can be expressed as:
\begin{equation}
m\left(i,j\right)_{t+f} = m^F\left(i,j\right)_{t+f|t} + \epsilon\left(i,j\right)_{t+f} ,
\end{equation}
where $m\left(i,j\right)_{t+f}$ is the \textit{observed} transition probability $(i,j)$ in period $t+f$, $m^F\left(i,j\right)_{t+f|t}$ is the \textit{forecasted} transition probability for the period $t+f$ calculated in period $t$ and $\epsilon\left(i,j\right)_{t+f}$ is the forecasting error. If the forecast is computed exploiting all the information available in period $t$, denoted by $\Omega_t$, then the expected value of $\epsilon\left(i,j\right)_{t+f}$ is zero and $\epsilon\left(i,j\right)_{t+f}$ and $m^F\left(i,j\right)_{t+f|t}$ are orthogonal, i.e.:
\begin{equation}
E[m\left(i,j\right)_{t+f} - m^F\left(i,j\right)_{t+f|t}|\Omega_t] = 0.
\end{equation}
Any \textit{significant} divergence between $m\left(i,j\right)_{t+f}$ and $m^F\left(i,j\right)_{t+f|t}$ signals a \textit{novelty} with respect to the information set available in period $t$, $\Omega_t$ or, in other words, $m^F\left(i,j\right)_{t+f|t}$ can be interpreted as a \textit{counterfactual} with respect to the event which happened at time $t$.
Hence, the effect of the introduction of a policy reform in period $t^\star$ should appear as a \textit{significant divergence} between $m\left(i,j\right)_{t^\star+f}$ and $m^F\left(i,j\right)_{t^\star+f|t^\star} \, \forall f \geq 1$.\footnote{If the policy reform  needs some time for its effects to be fully displayed, $f$ should be sufficiently long; however, the need to exclude other significant event in the period of forecast suggests to limit the length of $f$.}

In summary, to estimate the causal effects of a policy reform introduced at time $t^*$  in which all units are simultaneously treated after $t^*$, we need to follow a three-step process: (i) estimate the ARIMA model in the pre-intervention period; (ii) based on (i), perform a forecast to obtain an estimate of the counterfactual outcome in the post-intervention period, in the absence of intervention; (iii) to estimate the point causal effects as the difference between the observations and the corresponding forecasts at any  point in time in the post-intervention period. To perform inference on the estimated causal effects, bootstrap methods provide a natural way to calculate the empirical distribution of the estimates. They can also be used to conduct tests of hypothesis on the equality between $\mathbf{M}_{t^*+f:t^*+1}$ and $\mathbf{\widehat{M}}^F_{t^*+f:t^*+1|t}$ and between $\boldsymbol{\pi}_{t+f}$ and $\boldsymbol{\hat{\pi}}^F_{t+f|t}$, whose results will be crucial to evaluate the \textit{effectiveness} of the policy reform under scrutiny.

\section{The 2018 labour market reform \textit{Decreto Dignit\`a}}\label{sec:empirics}

In this section we contextualize the \textit{Decreto Dignit\`a} reform, which was implemented in Italy in 2018 to reduce job insecurity. We first describe the main features of the Italian labour market (Section  \ref{sec:italianLabourMarket}); then, we illustrate the  changes introduced by the \textit{Decreto Dignit\`a} (Section \ref{sec:DecretoDignità}), and the timing of the reform, which appears to be \textit{exogenous} and with no significant \textit{anticipation effects}  (Section \ref{sec:DecretoDignitàtiming}).

\subsection{The Italian labour market}\label{sec:italianLabourMarket}

Since 1990s labour market outcomes have improved substantially in Italy: \textit{employment} and \textit{labour force participation} have increased, and the \textit{unemployment rate} has dropped. But despite these improvements, the Italian labour market is still under-performing compared to those in most other European countries \citep{OECDreport2019}. Specifically, the participation rate is still substantially below that in most other European countries, the unemployment rate is higher, and the \textit{shares of temporary employment} and self-employment are significantly higher compared to the EU average (Table \ref{SEandTC}). The fast growing share of temporary employment since the 90s and the ambition to improve such labour market statistics led policymakers to implement several reforms over the years. Specifically, in March 2014 a labour market reform (\textit{Decreto Poletti}) reduced the rigidity of the employment protection legislation for temporary contracts to increase flexibility and stimulate job creation \citep{di2020heterogeneous}; in March 2015 the \textit{Jobs Act} changed the regulations of the open-ended contract, by introducing firing costs increasing with tenure \citep{boeri2019tale}; and, finally, in July 2018 the \textit{Decreto Dignit\`a} increased the rigidity of temporary contracts.
While these reforms aimed at targeting the way firms utilize temporary and permanent contracts, they might have also affected the unemployment and inactivity probabilities and the share of self-employment. In fact, they might have created incentives for individuals to enter or \textit{exit the labour force} and might have affected the probability to find a job or to be laid off. Moreover, the category of \textit{para-subordinate workers} in Italy, i.e. individuals who are legally self-employed, but often “economically dependent” on a single employer, which was  relatively large at the times of the reforms might have also been affected \citep{Raitano2018}.
In summary, the analysis suggests that five labour market states are relevant to assess the impact of the \textit{Decreto Dignit\`a}: inactivity, unemployment, temporary employment, permanent employment and self-employment.

\begin{table}[!htbp]
	\centering
	\caption{Labour market characteristics for a select sample of European countries. } 
	\label{SEandTC}
	\scriptsize
	\begin{tabular}{@{\extracolsep{-1pt}}lcccc} 
		\hline
		\hline
		\\[-1.8ex]
		\textbf{Country}&\textbf{Self-employment}&\textbf{Temporary-employment}&\textbf{Unemployment}&\textbf{Labour force participation}\\
		&(\% total employment)&(\% dependent employment)&(\% labour force)&(\% working age)\\
		\hline
		\\[-1.8ex]
		Greece&31.9&12.5&17.5&68.4\\
		\textbf{Italy}&	\textbf{22.7}&	\textbf{17.0}&	\textbf{10.2}&	\textbf{65.7}\\
		Portugal&16.9&20.8&6.7&75.5\\
		Spain&15.7&26.3&14.2&75.0\\
		United Kingdom&15.6&5.2&4.0&78.8\\
		Ireland&14.4&9.8&4.5&73.1\\
		Belgium&14.3&10.9&5.4&69.0\\
		France&12.1&16.4&8.5&71.7\\
		Germany&9.6&12.0&3.2&79.2\\\\[-1.8ex]
		\hline \\[-1.8ex]
		\textbf{EU average}&	\textbf{15.3}&	\textbf{13.2}&	\textbf{6.4}&	\textbf{74.2}\\
		\hline \hline
		\\[-1.8ex]
		\multicolumn{2}{l}{\textit{Source}: OECD, 2019.}
	\end{tabular} 
	\global\let\\=\restorecr
\end{table}

\subsection{The content of the reform \label{sec:DecretoDignità}}

The \textit{Decreto Dignit\`a},\footnote{Decree July 12, 2018, n. 87 converted into Law August 9, 2018, n. 96.} which was approved in July 2018, significantly increased the rigidity of the temporary contract legislation with the goal of reducing \textit{job instability}, defined as the time spent by individuals in \textit{temporary employment}. Specifically, the reform reduced the maximum length  of temporary contracts from 36 to 24 months. It also introduced the restriction that any temporary contract longer than 12 months could be utilized only in three circumstances: (i) to replace a worker, (ii) for temporary reasons, outside the regular business and (iii) in case of a temporary and unforeseeable increase in business. If the contract was not justified by any of these reasons, the contract would be transformed into a permanent one. The number of extensions within the 24 months was reduced from 5 to 4, and any renewal of the contract would need to be justified by any of the three reasons listed above. The reform also increased the social security contributions payable by employers for each temporary contract.\footnote{Prior to the reform coming into force, this contribution was set at 1.4\% of taxable salary for social security purposes and applied to all temporary contracts. With the reform it increased by 0.5\%. Moreover, the reform increased the firing costs associated with permanent contracts in case of unfair dismissals.} Clearly, the reform made the utilization of the temporary contract more difficult, more costly, it restricted the circumstances in which it could be utilized and it reduced the possibility of renewals/extensions.
As a result, some economists debated on the possible consequences of such reform, fearing that the increase in labour costs would lead to a decrease in labour demand and therefore an increase in unemployment, as predicted by the economic theory.\footnote{\href{https://www.lastampa.it/politica/2018/07/19/news/decreto-dignita-l-audizione-di-tito-boeri-la-stima-di-8-mila-posti-persi-e-ottimistica-1.34032792}{https://www.lastampa.it/politica/2018/07/19/news/+}.}

At the same time, the 2017 Budget Law, for the purpose of increasing permanent employment, introduced a norm by which employers hiring individuals below the age of 35 on a permanent contract in 2018 were entitled to a reduction of 50\% of the payable social security contributions for a maximum of 36 months with a cap of 3000\euro  \ annually. To be eligible employees should not have been hired ever before on a permanent contract. These incentives were then confirmed through the \textit{Decreto Dignit\`a} also in 2019 and 2020.

\subsection{The timing of the  reform \label{sec:DecretoDignitàtiming}}

In this section, we discuss the timing of the implementation of the 2018 reform to rule out the \textit{possibility of anticipation effects} and to provide evidence in support of the \textit{exogeneity of the reform} with respect to the labour market trends. Political elections took place in Italy on March 4, 2018. In the polls, the centre-right coalition, which included four political parties,\footnote{The centre-right coalition included: the League (right populist), Brothers of Italy (nationalist), Forza Italia (conservative) and the Us with Italy (Christian democrats).}  was listed as the most favourite, followed by the populist Five Star Movement (M5S) and the centre-left coalition.\footnote{The centre-left coalition included four parties: Democratic Party, More Europe (liberal), Together (progressive), Popular Civic list (Christian democrats)).}  None of those parties  had listed plans on implementing labour market reforms to reduce job uncertainty and support the transitions from temporary to permanent employment in their political manifesto. 

The outcome of the election was that the centre-right coalition emerged with a plurality of seats in the Chamber of Deputies and in the Senate, while the anti-establishment Five Star Movement became the party with the largest number of votes. The centre-left coalition came third. As no political group or party won an outright majority, the election resulted in a hung parliament. The institutional crisis lasted for 3 months, until 21 May, when unexpectedly, the M5S and one of the parties in the centre-right coalition, the League, reached an agreement on a government program, clearing the way for the formation of a governing coalition between the two parties and proposing law professor Giuseppe Conte as Prime Minister. The Government was officially formed on June 1, 2018. Immediately after its constitution, the Ministry of Labour, Luigi Di Maio, declared his intention to improve the job security of Italian people, by regulating the gig economy and making temporary employment more costly for employers. He mentioned in several public media speeches that his mission was to eliminate precarious employment, i.e., the persistence of workers into temporary employment. The \textit{Decreto Dignit\`a} was then first presented as a legislative decree in the Parliament on July 12, 2018 and converted into law on August 9, 2018.

Given the absence of any mention of such reform in the party program, the unexpected coalition between the two parties which formed the Government, and the very short time (one month) between the settlement of the new Government and the presentation of the first version of the reform, we argue that the possibility of any significant anticipation effect is minimal and the reform can be considered as an exogenous shock to the economic dynamics. Finally, no other significant reform and/or interventions in the labour market took place until the outburst of pandemic crisis in 2020.

\section{Data \label{sec:dataset}}

We use Italian quarterly longitudinal labour force data as provided by the Italian Institute of Statistics (ISTAT) for the period 2013 (quarter I) to 2020 (quarter IV).\footnote{Data for the period 2013 (quarter I) to 2019 (quarter IV) are available upon request at: https://www.istat.it/it/archivio/185540.} The Italian Labour Force Survey (LFS) follows a simple rotating sample design where households participate for two consecutive quarters, exit for the following two quarters, and come back in the sample for other two consecutive quarters. As a result, 50\% of the households, interviewed in a quarter, are re-interviewed after three months, 50\% after twelve months, 25\% after nine and fifteen months. This rotation scheme allows to obtain 3 months longitudinal data, which include almost 50\% of the original sample.

The longitudinal feature of these data is essential for achieving a complete picture of significant economic phenomena of labour market mobility. Per each individual who has been interviewed, we observe a large number of individual and labour market characteristics at the time of the interview and three months before. Taking into account the structure of this database, we compute the labour market flows by calculating the quarter-on-quarter transitions made by individuals between different labour market states. Specifically, we estimate the gross flows using a five-state model (permanent employment, temporary employment, self-employment, unemployment, and inactivity).
On average approximately 70.000 individuals are interviewed each quarter, of which 45.000 are part of the working age population. The average quarterly inflow of younger individuals in the working age population is 0.3\%, while the average quarterly outflow of older individuals from the working age population is 0.4\%, backing our hypothesis of a (almost) constant working age population within quarters (Section \ref{sec:methodology}).

\section{An empirical evaluation of Decreto Dignit\`a} \label{sec:evaluationDecretoDignità}

In this section we first discuss the validity of our approach, based on the fulfillment of the assumptions described in Section \ref{sec:counterfactualViaARIMA}. We then compare the forecasted and observed path of labour shares and transitional probabilities in the first four quarters after the introduction of the reform (Sections \ref{sec:labourMarketShares} and  \ref{sec:transtitionProbabilities}). Section \ref{sec:quantitative} computes the cumulative effects of the reform and their statistical significance. Finally, Section \ref{sec:heterogeneousEffects} explores heterogeneous effects for specific categories of workers (by age, gender,education, etc.).

\subsection{The casual evaluation of Decreto Dignit\`a \label{sec:casualEvaluationDecreto}}

The C-ARIMA approach described in Section \ref{sec:counterfactualViaARIMA} is suitable to estimate the causal effect  of the \textit{Decreto Dignit\`a}, given its features. First, the policy reform affected all individuals \textit{erga omnes} and its implementation was perceived by workers and firms as permanent (Section \ref{sec:counterfactualViaARIMA}), thus  satisfying the assumptions of single persistent intervention for all units at the same time and no temporal interference. Moreover, its introduction was unexpected, thus satisfying the no anticipatory effects assumption, as explained in Section \ref{sec:DecretoDignitàtiming}.
The forecast will be calculated by estimating an ARIMA model with no covariates and within a time horizon of one year to comply with the assumption of conditional stationarity of the model. On the basis of all the assumptions of Section \ref{sec:counterfactualViaARIMA} being fulfilled, we claim that the comparison between the forecasted and the observed labour shares and transitional probabilities causally quantifies the effect of the \textit{ Decreto Dignit\`a}.
        
We take  quarter III of 2018, which includes the \textit{date of the approval} of the \textit{Decreto Dignit\`a} (July 12, 2018), as the time when the dynamics of the Italian labour market is expected to change. A change in the shares usually takes more time to materialize as it is the result of cumulated changes in transition probabilities. Moreover, changes in the shares could hide compositional dynamics, which are observable only when looking at the flows of individuals across labour market states. Thus, to fully understand the effect of the \textit{Decreto Dignit\`a}, we will evaluate the changes both in the \textit{shares of workers} in each labour market state and in the \textit{transition probabilities} between labour market states.

\subsection{Labour market shares \label{sec:labourMarketShares}}

We first look at the evolution of the five labour market shares in each quarter from 2013 (quarter II) to 2019 (quarter IV). 
The objective of this exercise is two-fold: first, we aim to identify the best time period on which to focus our analysis on; second, we aim to assess the casual effect of the reform on the shares.

\begin{figure}[!htbp]
	\centering
	\caption{Observed shares of individuals in different labour market states (\% of working age population).}
	\label{fig:ObsMasses}
	\begin{subfigure}[b]{0.3\textwidth}
		\centering
		\includegraphics[width=\linewidth]{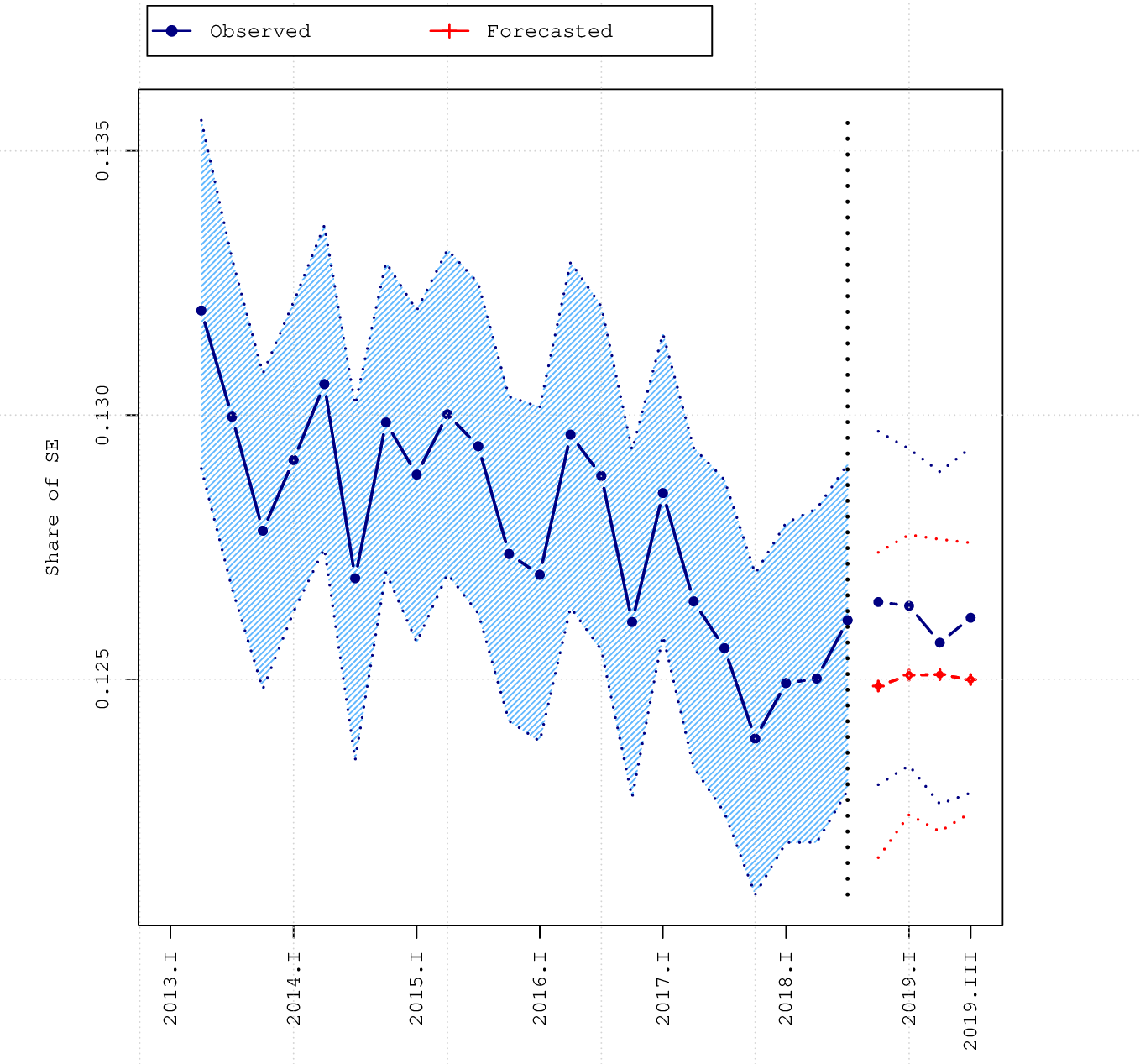}
		\caption{Self-employed.}
		\label{shareSEAll}
	\end{subfigure}
	\begin{subfigure}[b]{0.3\textwidth}
		\centering
		\includegraphics[width=\linewidth]{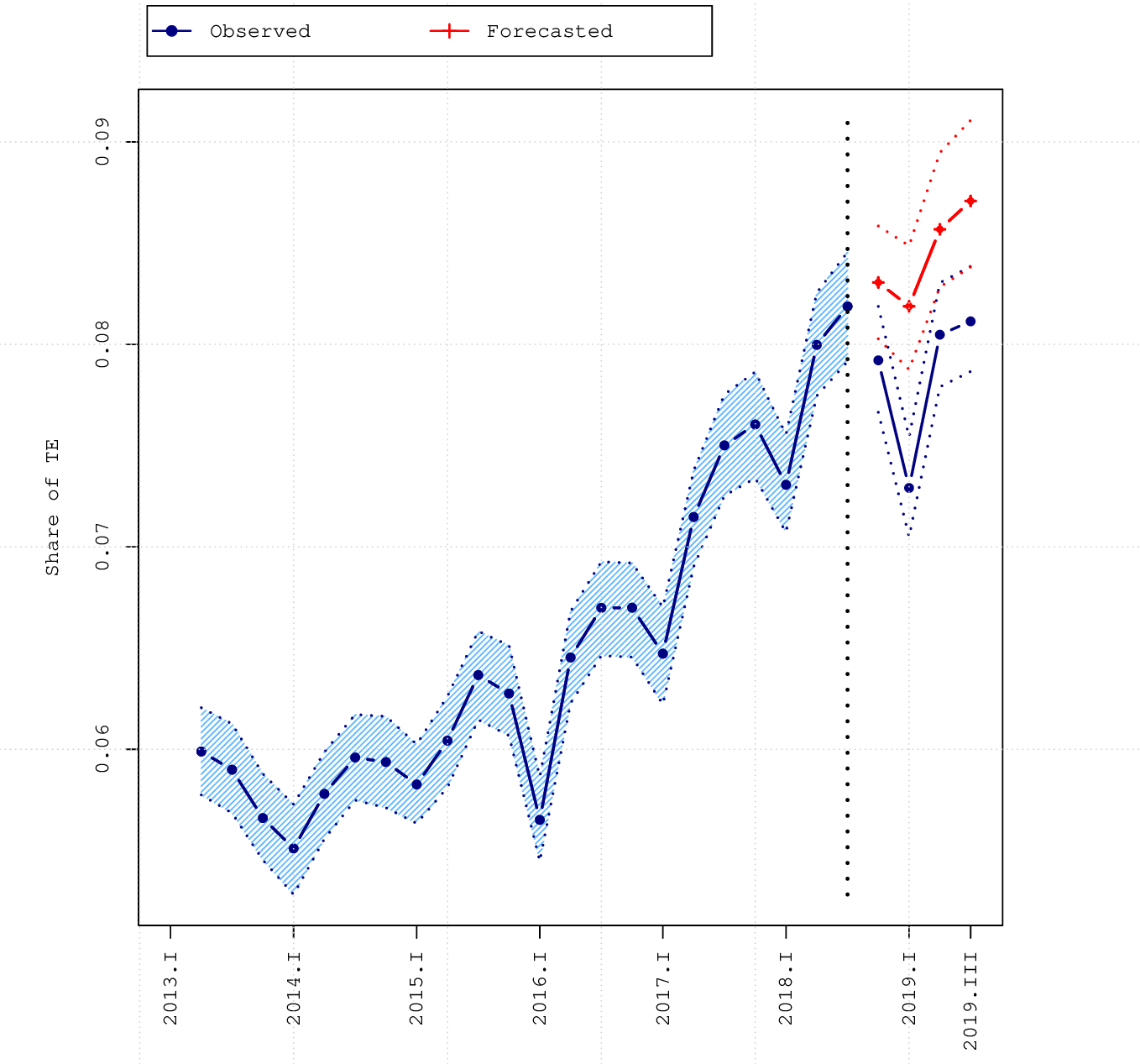}
		\caption{Temporary employed.}
		\label{shareTEAll}
	\end{subfigure}
	\begin{subfigure}[b]{0.3\textwidth}
		\centering
		\includegraphics[width=\linewidth]{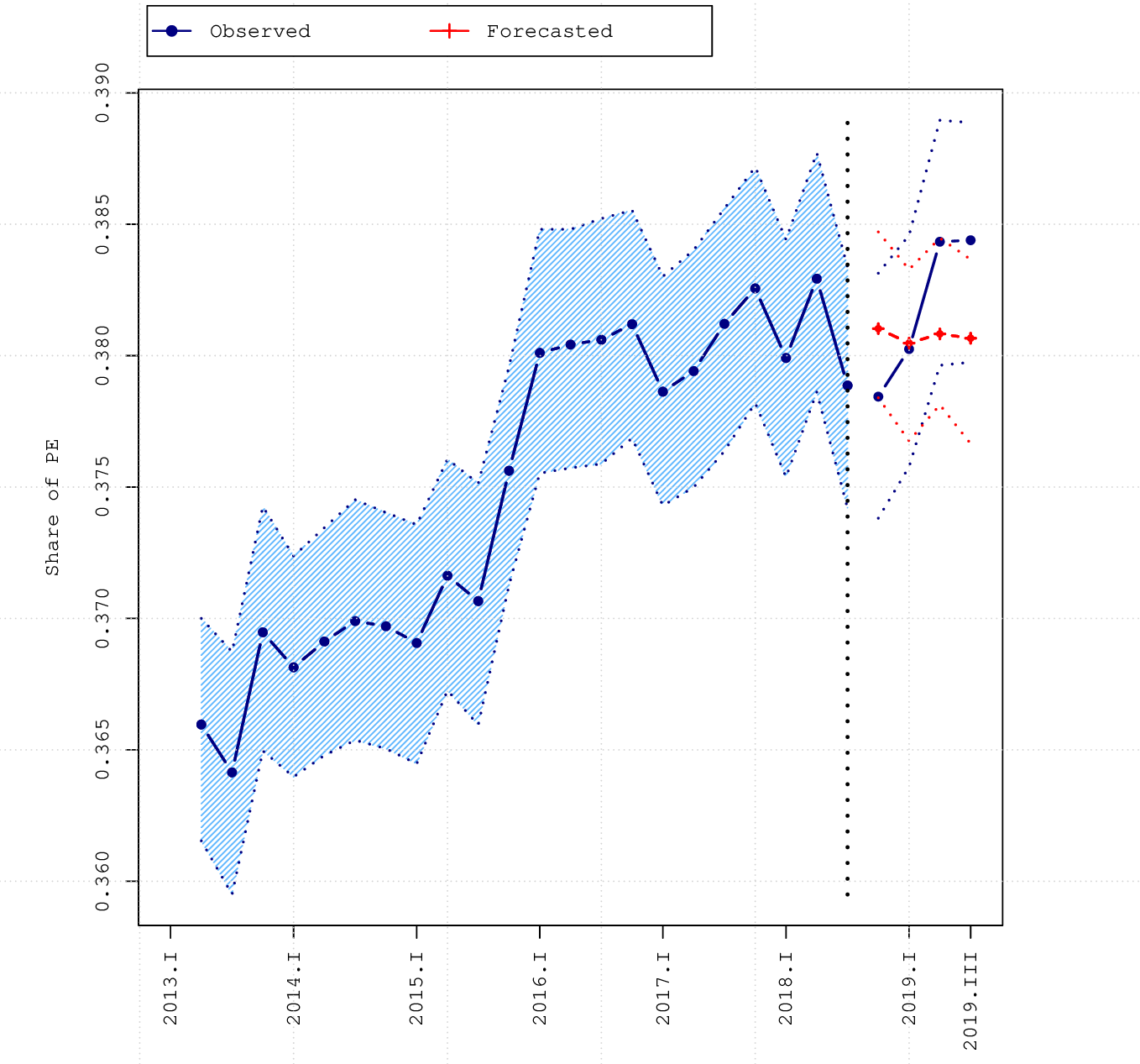}
		\caption{Permanent employed.}
		\label{sharePEAll}
	\end{subfigure}\\
	\begin{subfigure}[b]{0.3\textwidth}
		\centering
		\includegraphics[width=\linewidth]{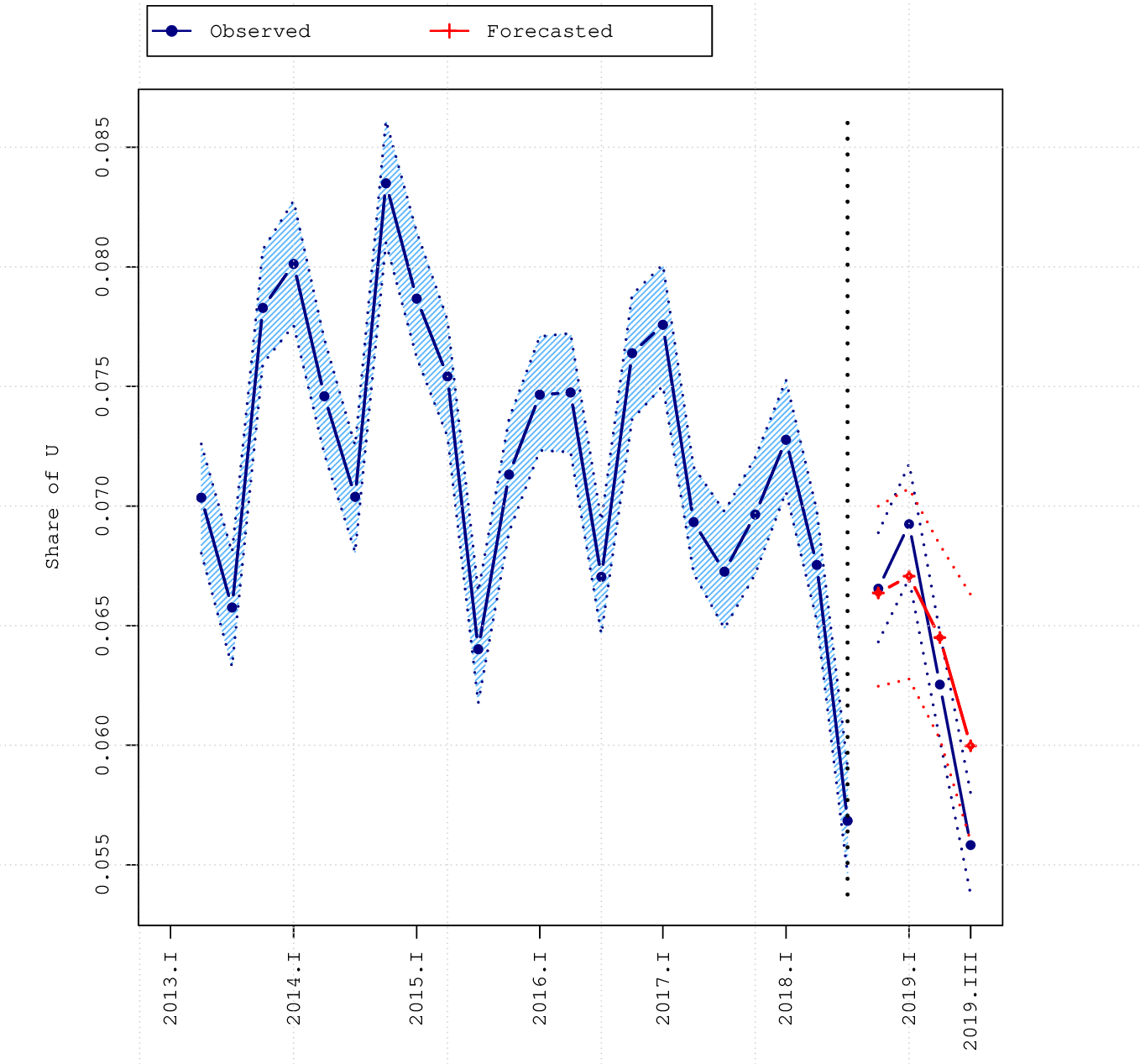}
		\caption{Unemployed.}
		\label{shareUAll}
	\end{subfigure}	
	\begin{subfigure}[b]{0.3\textwidth}
		\centering
		\includegraphics[width=\linewidth]{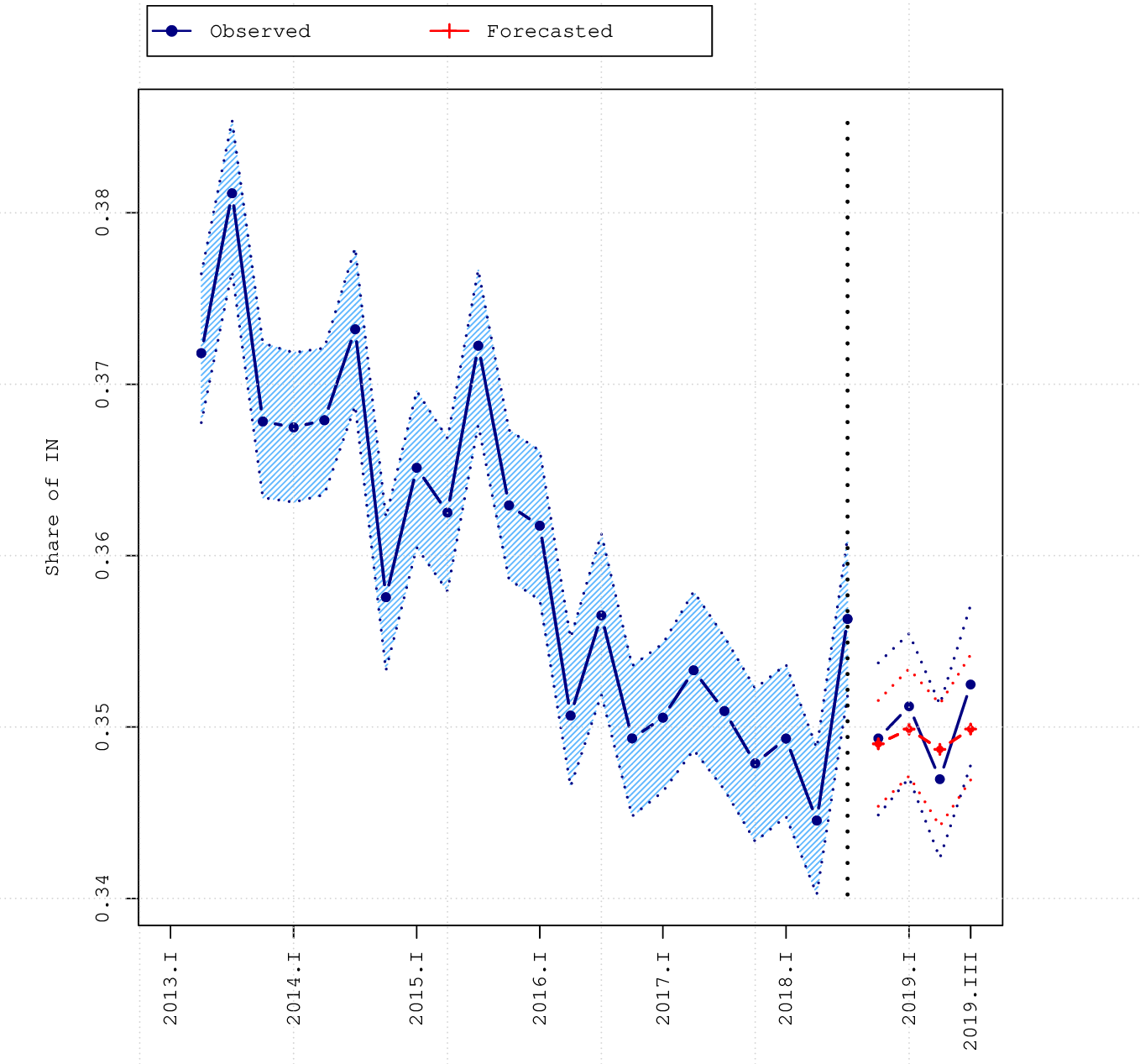}
		\caption{Inactive individuals.}
		\label{shareINAll}
	\end{subfigure}
	\caption*{ \scriptsize{Note: the blue line reports the observed share of individuals in each labour market state, while the red line reports the counterfactual share, with 95\% confidence intervals calculated via bootstrap (see Appendix \ref{app:bootstrapProcedures}). The vertical dotted line represents the \textit{Decreto Dignit\`a} implemented in July 2018. }}
\end{figure}

 The observed share of individuals hired on a \textit{temporary contract} was stable in the period 2013-2016. The share has been then increasing since 2016, until the \textit{Decreto Dignit\`a}  interrupted the positive trend. The share remained then stable until the end of 2019.  The share of \textit{permanent employees} was also stable around 37\% between between 2013 and 2015, when it jumped to 38\%, and remained then stable between 2016 and the end of 2019.
 We also observe a declining trend in the \textit{share of inactive individuals} from 2013, which slows down in 2016. Since then we observe a rather flat pattern. In terms of \textit{unemployment and self-employment}, the shares seem to be constant between 2013 and 2016, while showing a declining trend afterwards. 
 These figures suggest that the period 2016-2019 is the one we should focus our analysis on. This choice is also supported by the fact that a number of labour market reforms took place in 2014 and 2015 (Section \ref{sec:italianLabourMarket}), which could confound our estimates. Thus, we base our forecasts on the observation period which runs from quarter I of 2016 to quarter III of 2018. From quarter IV of 2018, we report together with the annual observed shares (blue line) the forecasted shares (red line), which we take as the counterfactuals without \textit{Decreto Dignit\`a}, together with their 95\% confidence intervals, calculated via bootstrap (see Appendix \ref{app:bootstrapProcedures}).
 
 We observe that the share of individuals who were employed on a temporary contract would have reached approximately 8.9\% of the working age population after one year if no reform would have been implemented, against the observed average share of 8.1\%, with this difference being statistically significant. We do not report any other significant difference between the observed and forecasted shares across the other labour market states.

 \subsection{Transition probabilities \label{sec:transtitionProbabilities}}

\begin{figure}[!htbp]
 	\centering
 	\caption{Transition probabilities across labour market states.}
 	\label{fig:transIntensities1}
 	\begin{subfigure}[b]{0.32\textwidth}
 		\centering
 		\includegraphics[width=\linewidth]{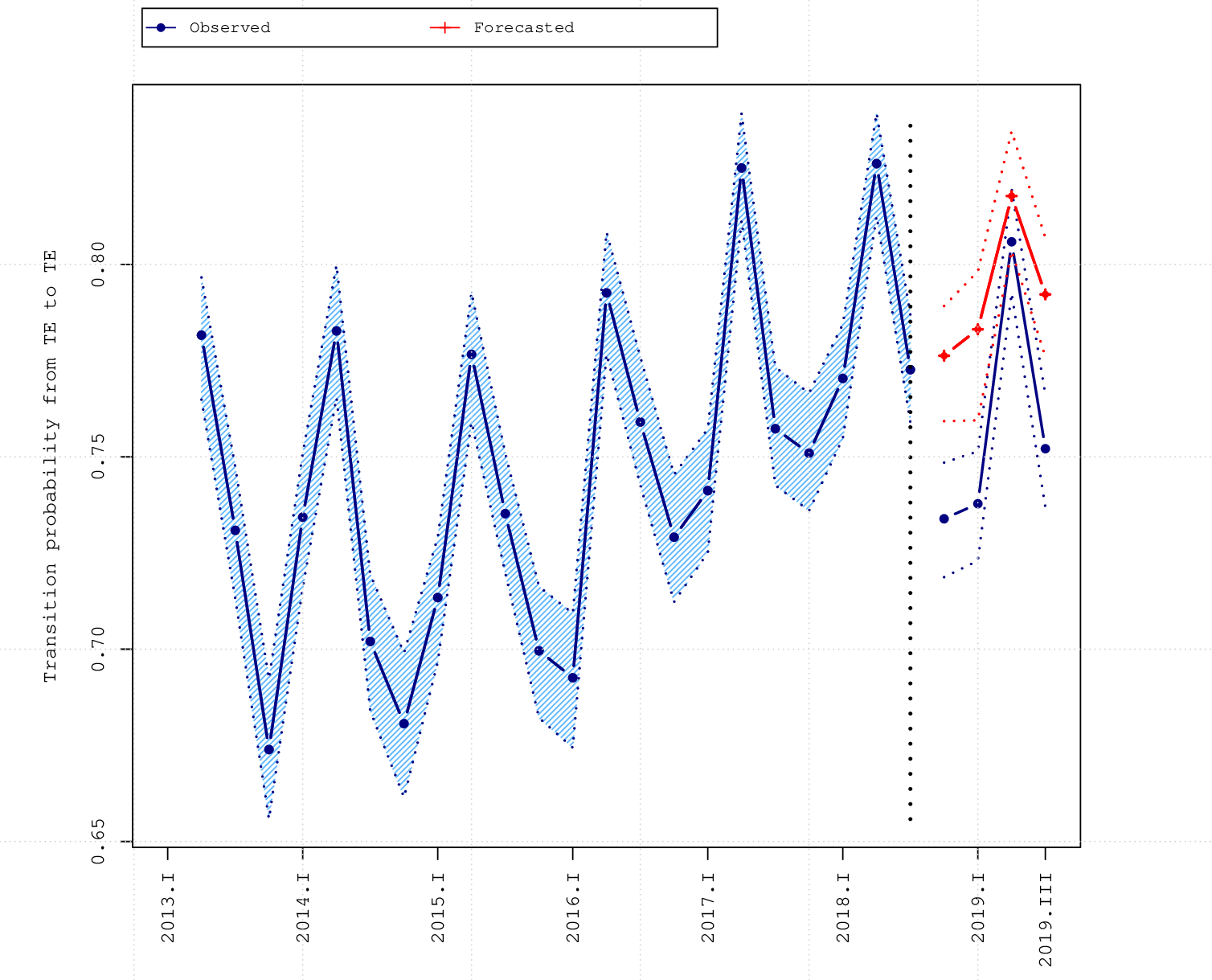}
 		\caption{From TE to TE.}
 		\label{transProbFromTEtoTEAll}
 	\end{subfigure}
 	\begin{subfigure}[b]{0.32\textwidth}
 		\centering
 		\includegraphics[width=\linewidth]{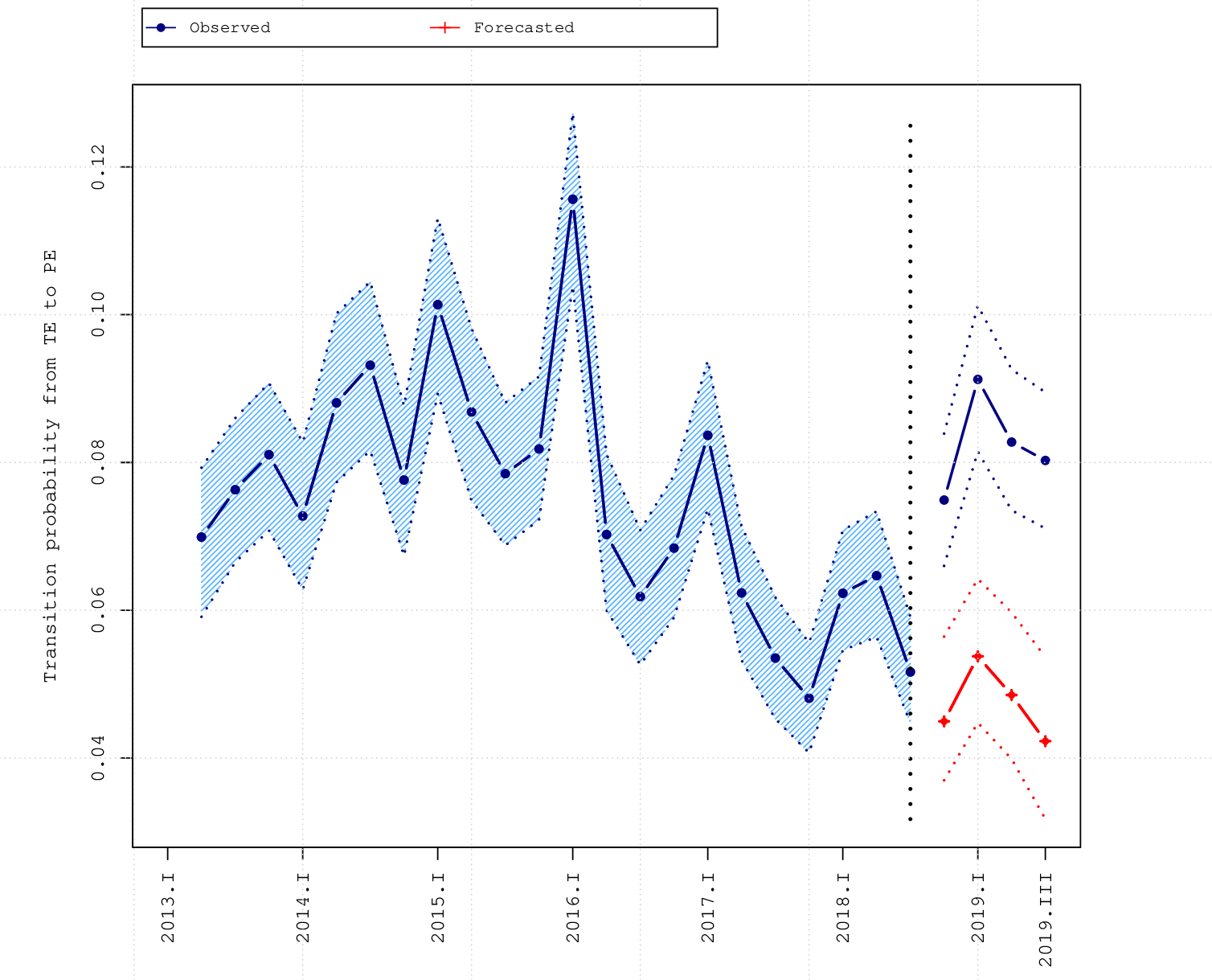}
 		\caption{From TE to PE.}
 		\label{transProbFromTEtoPEAll}
 	\end{subfigure}
 	\begin{subfigure}[b]{0.32\textwidth}
 		\centering
 		\includegraphics[width=\linewidth]{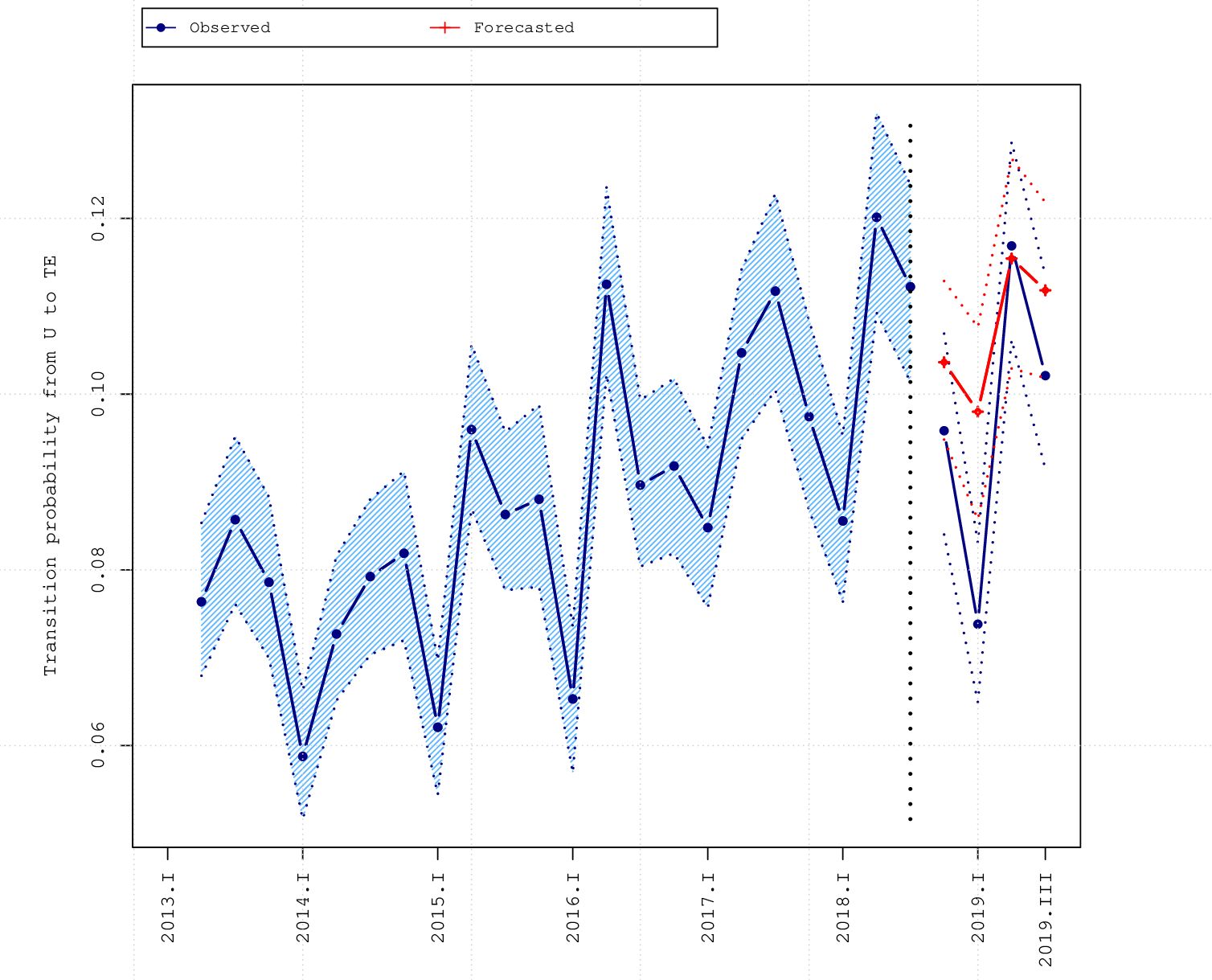}
 		\caption{From U to TE.}
 		\label{transProbFromUtoTEAll}
 	\end{subfigure}
 	\begin{subfigure}[b]{0.32\textwidth}
 		\centering
 		\includegraphics[width=\linewidth]{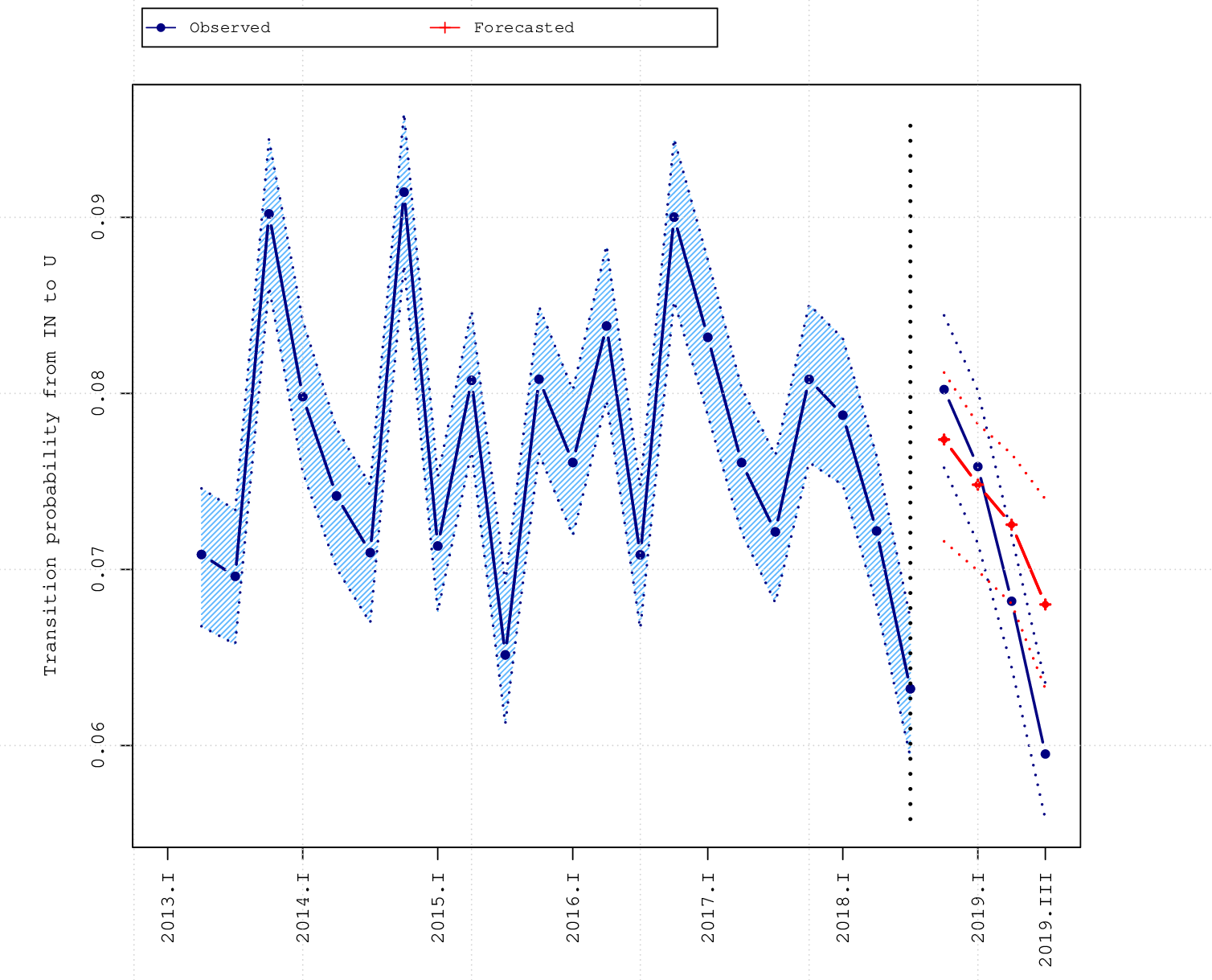}
 		\caption{From IN to U.}
 		\label{transProbFromINtoUAll}
 	\end{subfigure}
 	\begin{subfigure}[b]{0.32\textwidth}
 		\centering
 		\includegraphics[width=\linewidth]{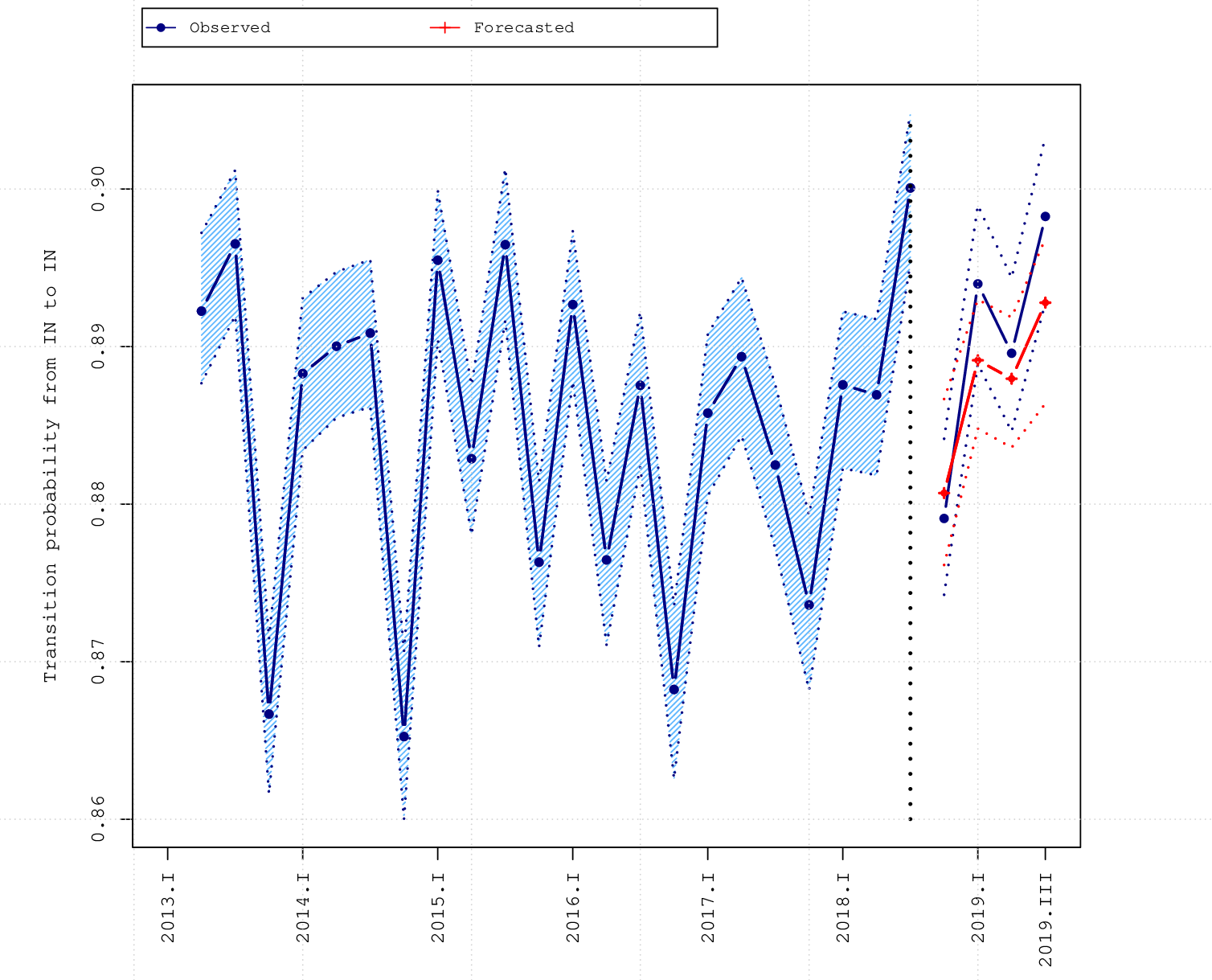}
 		\caption{From IN to IN.}
 		\label{transProbFromINtoINAll}
 	\end{subfigure}
 	\caption*{ \scriptsize{Note: the blue line reports the estimated transition probabilities, while the red line reports the counterfactual transition probabilities, with 95\% confidence intervals calculated via bootstrap (see Appendix \ref{app:bootstrapProcedures}). The vertical dotted line represents the \textit{Decreto Dignit\`a}  implemented in July 2018. }}
 \end{figure}

 To dig deeper into the underlying dynamics, we document how the transition probabilities across the five labour market states considered  evolved  from 2013 (quarter I) to 2019 (quarter IV). Specifically, the transition probabilities which we believe are relevant for explaining the changes in the shares highlighted above are reported in Figure \ref{fig:transIntensities1}. 
 
\textit{Persistence in temporary employment} and transitions from \textit{temporary to permanent employment} changed remarkably  after 2016. Specifically, persistence into temporary employment was fluctuating around 73\% in the period 2013-2016. Since then it increased to almost 80\%. The transitions from temporary  to permanent employment which were stable around 80\% before 2016,  significantly decreased afterwards.
Transition probabilities from  unemployment to temporary employment were also stable before 2016,  but increased afterwards. Opposite patterns are found for the transitions from inactivity to unemployment: stable before 2016, and declining later on. Persistence  in the inactivity state seems to have declined after 2016, while being stable in the period 2013-2016.
Overall, these figures point to a change of patterns after 2016, suggesting once again the period 2016-2019 to be the most suitable for our analysis.
When looking at the changes in the transition probabilities after the 2018 reform, we observe a \textit{larger flow of individuals upgrading from temporary to permanent employment}. This evidence is a first indicator that the objectives of the reform might have been met. We do not find from these statistics any concerning signals regarding the outflows: no significant changes are reported in the outflows \textit{from unemployment towards temporary employment}, and \textit{from inactivity towards unemployment}. Moreover, no significant differences between the observed and counterfactual probabilities are reported in relation to the \textit{persistence in inactivity}.

\subsection{Quantifying the cumulative impact of the reform \label{sec:quantitative}}

So far, we have provided evidence about the changes in the labour market dynamics following the implementation of the 2018 reform. In this section, we will quantify those changes and provide some statistical tests.
To do so, we first compute the \textit{fitted shares} of individuals in the five labour market states in the four periods following the reform, i.e., from quarter IV of 2018 to quarter III of 2019.\footnote{Note that the observed shares are slightly different compared to the fitted shares as the assumption underneath our estimates are of a constant working age population, while in the data we observe that in the period considered the working age population declines. This is likely due to population aging, although mortality and migration might have played a minor role as well.} Following Section \ref{sec:counterfactualViaARIMA}, these shares are computed by multiplying the estimated transition matrices per each quarter in the period considered with the observed shares in the quarter in which the reform was implemented, i.e., quarter III of 2018. 
We then repeat the same exercise using the forecasted transition matrices to get the \textit{forecasted shares} of individuals in the five labour market states in the four periods following the reform. The differences between fitted and forecasted shares represent our measure of the impact of the reform.

Table \ref{tab:FittedForecastedShares} shows significantly higher fitted shares compared to forecasted ones in permanent employment (+1.1 pp) and significantly lower shares in temporary employment (-1 pp) and unemployment (-0.5 pp). We do not find significant changes in the shares of self-employed and inactive.

\begin{table}[!htbp]
	\scriptsize
	\caption{Fitted versus forecasted statistics.}
	\centering
	\begin{subtable}[h]{\textwidth}
	\caption{Fitted versus forecasted  shares of individuals in different labour market states from quarter IV of 2018 to quarter III of 2019.}
	\label{tab:FittedForecastedShares}
	\centering
	\begin{tabular}{r|ccccc}
		\\[-1.8ex]\hline 
	\hline \\[-1.8ex] 
	& SE & TE & PE & U & IN \\ 
	\hline \\[-1.8ex]
Fitted & $0.125$ & $\textbf{0.080}$ & $\textbf{0.380}$ & $\textbf{0.054}$ & $0.361$\\
(s.e.) & (0.001) & (0.001) & ($0.001$) & ($0.001$) & ($0.002$) \\ 
Forecasted & $0.123$ & $\textbf{0.090}$ & $\textbf{0.369}$ & $\textbf{0.059}$ & $0.359$ \\ 
(s.e.) & ($0.001$) & ($0.002$) & ($0.002$) & ($0.001$) & ($0.002$) \\ 
\hline
\end{tabular} 
	\end{subtable}\\
\vspace{0.5cm}

\begin{subtable}[!htbp]{\textwidth}
	\caption{Difference between fitted and forecasted numbers of individuals in different labour market states from quarter IV of 2018 to quarter IV of 2019.}
	\label{tab:FittedForecastedNumbers}
	\centering
	\begin{tabular}{r|rrrrr}
	\\[-1.8ex]\hline 
\hline \\[-1.8ex] 
& SE & TE & PE & U & IN \\ 
\hline \\[-1.8ex]
		C.I. 97.5\% & $154,685$ & $$-$196,169$ & $548,449$ & $$-$13,979$ & $290,559$ \\ 
Difference & $74,275$ & $$-$\textbf{379,650}$ & $\textbf{407,857}$ & $$-$\textbf{172,624}$ & $70,142$ \\ 
C.I. 2.5\% & $$-$80,843$ & $$-$519,063$ & $218,816$ & $$-$289,958$ & $$-$112,291$ \\   
		\hline
	\end{tabular}\\
\caption*{\textit{Note}: Confidence intervals are calculated via bootstrap (see Appendix \ref{app:bootstrapProcedures}).
}
\end{subtable}\\
\vspace{0.5cm}
\begin{subtable}[!htbp]{\textwidth}
	\caption{Fitted versus forecasted cumulative transition probabilities from quarter IV of 2018 to quarter IV of 2019.}
	\label{tab:FittedForecastedTransProb}
	\centering
	\begin{tabular}{r|rrrrr}
	\\[-1.8ex]\hline 
\hline \\[-1.8ex] 
& SE & TE & PE & U & IN \\ 
\hline \\[-1.8ex]
		SE & $0.014$ & $$-$\textbf{0.005}$ & $$-$0.001$ & $$-$0.003$ & $$-$0.005$ \\ 
TE & $$-$0.001$ & $$-$\textbf{0.066}$ & $\textbf{0.081}$ & $$-$0.005$ & $$-$0.009$ \\ 
PE & $0.001$ & $$-$\textbf{0.003}$ & $0.005$ & $$-$0.002$ & $$-$0.001$ \\ 
U & $$-$0.003$ & $$-$\textbf{0.019}$ & $\textbf{0.014}$ & $$-$0.006$ & $0.013$ \\ 
IN & $0$ & $$-$0.004$ & $0.004$ & $$-$\textbf{0.008}$ & $0.008$ \\ 
\hline    
	\end{tabular}\\
	\caption*{\textit{Note}: in bold we report probabilities, calculated via bootstrap (see Appendix \ref{app:bootstrapProcedures}), which are statistically significant at 5\% level.}
\end{subtable}
\end{table}

We  then multiply these differences in the shares by the number of working age individuals in Italy in quarter III of 2019 to quantify the magnitude of the impact of the reform (Table \ref{tab:FittedForecastedNumbers}). It is estimated that there were 379.000 fewer employees in temporary employment, 172,000 fewer in unemployment and about 407.000 more in permanent employment after the reform. 
Finally, we compute the difference between the fitted and forecasted cumulative transition probabilities for the four quarters after the reform to understand how the flows have responded to the reform (Table \ref{tab:FittedForecastedTransProb}). 
We confirm that the increased rigidity of temporary contracts  led to a reduced persistence in temporary employment (-6.6 pp) and increased flows from temporary to permanent employment (+8.1 pp). The additional rigidity also decreased the flows from unemployment to temporary employment (-1.9pp). We also observe less participation to the labour market: the flows from inactivity to unemployment  decreased (-0.8 pp) and the flows from unemployment to inactivity  increased (+1.3 pp), although the latter is not statistically significant at 5\% level. The decrease in unemployment is therefore the outcome of a \textit{composition effect} in the transition probabilities and not a direct effect of more inflows in unemployment and/or less outflows (see Appedix \ref{app:transitionBetweenStatus}).

Overall, we can conclude that the additional rigidity  favored the \textit{upgrading from temporary to permanent employment} and thus reduced job uncertainty, as advocated by the reform. Moreover, it reduced unemployment by redirecting unemployed workers  towards permanent contracts. We find also signs of decreased participation to the labour market.

Our results are robust to a small change in the treatment time ($t^*$) (Appendix \ref{app:Q32018}). Finally, our methodology passes a placebo test where $t^*$ is moved to quarter III of 2017 (Appendix \ref{app:placebo}).

\subsection{Heterogeneous effects \label{sec:heterogeneousEffects}} 

The literature points to some heterogeneity in the probability to be in temporary employment \citep{tealdi2019adverse, berton2015non}.  Thus, we  look at the shares of individuals in different labour market states by categories based on age, gender, education and geographical location. Specifically, we consider young those workers who are younger than 35, as this is the threshold for employers to be eligible for fiscal incentives. We define as high educated those individuals with a tertiary level of education and finally we split the observations by geographical area, considering the South of Italy versus the Center/North.

\begin{table}[!htbp]
	\scriptsize
	\caption{Fitted versus forecasted statistics - Females.}
	\begin{subtable}[h]{\textwidth}
		\caption{Fitted versus forecasted  shares of individuals in different labour market states from quarter IV of 2018 to quarter IV of 2019- Females.}
		\label{tab:FittedForecastedSharesFemales}
		\centering
		\begin{tabular}{@{\extracolsep{5pt}} cccccc} 
			\\[-1.8ex]\hline 
			\hline \\[-1.8ex] 
			& SE & TE & PE & U & IN \\ 
			\hline \\[-1.8ex]
			Fitted & $0.079$ & $\textbf{0.072}$ & $\textbf{0.339}$ & $0.049$ & $0.461$ \\  
			(s.e) & ($0.001$) & ($0.002$) & ($0.002$) & ($0.001$) & ($0.002$) \\ 
			Forecasted & $0.074$ & $\textbf{0.083}$ & $\textbf{0.327}$ & $0.053$ & $0.464$ \\  
			(s.e) & ($0.002$) & ($0.002$) & ($0.002$) & ($0.002$) & ($0.003$) \\  
			
			\hline 
		\end{tabular}
	\end{subtable}
	\\
	
		\vspace{0.5cm}
		\begin{subtable}[h]{\textwidth}
			\caption{Difference between fitted and forecasted numbers of individuals in different labour market states from quarter IV of 2018 to quarter IV of 2019 - Females.}
			\label{tab:FittedForecastedNumbersFemales}
			\centering
			\begin{tabular}{@{\extracolsep{5pt}} cccccc} 
				\\[-1.8ex]\hline 
				\hline \\[-1.8ex] 
				& SE & TE & PE & U & IN \\ 
				\hline \\[-1.8ex] 
				C.I. 97.5\% & $159,382$ & $$-$88,441$ & $315,576$ & $42,535$ & $90,551$ \\ 
				Difference & $93,995$ & $$-$\textbf{209,467}$ & $\textbf{225,384}$ & $$-$62,653$ & $$-$47,259$ \\ 
				C.I. 2.5\% & $3,621$ & $$-$296,672$ & $108,292$ & $$-$133,085$ & $$-$191,607$ \\    
				
				\hline \\[-1.8ex] 
			\end{tabular} 
			\vspace{0.3cm}
		\end{subtable} 
		\begin{subtable}[h]{\textwidth}
			\caption{Fitted versus forecasted cumulative transition probabilities from quarter IV of 2018 to quarter IV of 2019 - Females.}
			\label{tab:FittedForecastedTransProbFemales}
			\centering 
			\begin{tabular}{@{\extracolsep{5pt}} cccccc} 
				\\[-1.8ex]\hline 
				\hline \\[-1.8ex] 
				& SE & TE & PE & U & IN \\ 
				\hline \\[-1.8ex] 
				SE & $\textbf{0.031}$ & $$-$\textbf{0.012}$ & $$-$0.012$ & $$-$0.001$ & $$-$0.006$ \\ 
				TE & $0.008$ & $$-$\textbf{0.074}$ & $\textbf{0.105}$ & $$-$0.002$ & $$-$\textbf{0.037}$ \\ 
				PE & $0.002$ & $$-$0.003$ & $0.002$ & $$-$0.001$ & $0$ \\ 
				U & $$-$0.004$ & $$-$\textbf{0.015}$ & $\textbf{0.019}$ & $$-$0.003$ & $0.004$ \\ 
				IN & $0.003$ & $$-$0.006$ & $\textbf{0.007}$ & $$-$0.006$ & $0.001$ \\ 
				\hline \\[-1.8ex] 
			\end{tabular} \\
		\end{subtable} 
		\caption*{\scriptsize{\textit{Note}: in bold we report probabilities, calculated via bootstrap (see Appendix \ref{app:bootstrapProcedures}), which are statistically significant at 5\% level.}}
	\end{table}

\paragraph{Gender}

The dynamics of females have been quite different compared to the dynamics of males. From Table \ref{tab:FittedForecastedSharesFemales}, we observe a significant increase in the shares of permanent employed (1.2 pp) and a large decrease in the share of temporary employed (-1.1 pp), while the shares of self-employed, unemployed and inactive women are unchanged. These correspond to approximately 209,000 fewer women in temporary employment, 225,000 more in permanent employment (Table \ref{tab:FittedForecastedNumbersFemales}). Females previously on temporary contracts persist less in temporary employment (-7.4 pp), move less towards inactivity (-3.7 pp), and instead transit more towards permanent employment (+10.5 pp)(Table \ref{tab:FittedForecastedTransProbFemales}). The cumulative transition probabilities also point to a remarkable increase in the inflow to permanent employment from inactivity (+0.7 pp) and unemployment (+1.9 pp). Women seem to be great beneficiaries of the reform, as their job uncertainty is reduced, with positive  effects even in terms of labour force participation.

\begin{table}[!htbp]
	\scriptsize
	\caption{Fitted versus forecasted statistics - Males.}
	\begin{subtable}[h]{\textwidth}
		\caption{Fitted versus forecasted  shares of individuals in different labour market states from quarter IV of 2018 to quarter IV of 2019- Males.}
		\label{tab:FittedForecastedSharesMales}
		\centering
		\begin{tabular}{@{\extracolsep{5pt}} cccccc} 
			\\[-1.8ex]\hline 
			\hline \\[-1.8ex] 
			& SE & TE & PE & U & IN \\ 
			\hline \\[-1.8ex] 
			Fitted  & $0.171$ & $\textbf{0.089}$ & $\textbf{0.420}$ & $\textbf{0.059}$ & $\textbf{0.260}$ \\ 
			(s.e.)& ($0.001$) & ($0.002$) & ($0.002$) & ($0.002$) & ($0.002$) \\ 
			Forecasted & $0.172$ & $\textbf{0.096}$ & $\textbf{0.412}$ & $\textbf{0.068}$ & $\textbf{0.251}$ \\ 
			(s.e.) & ($0.002$) & ($0.002$) & ($0.002$ & ($0.002$) & ($0.002$) \\ 
			\hline 
		\end{tabular}
		\vspace{0.5cm}
	\end{subtable}
	
	\begin{subtable}[h]{\textwidth}
		\caption{Difference between fitted and forecasted numbers of individuals in different labour market states from quarter IV of 2018 to quarter IV of 2019 - Males.}
		\label{tab:FittedForecastedNumbersMales}
		\centering
		\begin{tabular}{@{\extracolsep{5pt}} cccccc} 
			\\[-1.8ex]\hline 
			\hline \\[-1.8ex] 
			& SE & TE & PE & U & IN \\ 
			\hline \\[-1.8ex] 
			C.I. 97.5\% & $42,104$ & $$-$56,790$ & $288,999$ & $$-$39,341$ & $287,725$ \\ 
			Difference & $$-$30,525$ & $$-$\textbf{174,799}$ & $\textbf{174,754}$ & $$-$\textbf{128,107}$ & $\textbf{158,676}$ \\ 
			C.I. 2.5\% & $$-$115,050$ & $$-$272,447$ & $47,594$ & $$-$219,287$ & $31,870$ \\  
			\hline \\[-1.8ex] 
		\end{tabular} 
		\vspace{0.3cm}
	\end{subtable} 
	\vspace{0.5cm}
	\begin{subtable}[h]{\textwidth}
		\caption{Fitted versus forecasted cumulative transition probabilities from quarter IV of 2018 to quarter IV of 2019 - Males.}
		\label{tab:FittedForecastedTransProbMales}
		\centering 
		\begin{tabular}{@{\extracolsep{5pt}} cccccc} 
			\\[-1.8ex]\hline 
			\hline \\[-1.8ex] 
			& SE & TE & PE & U & IN \\ 
			\hline \\[-1.8ex] 
			SE & $0.005$ & $$-$0.002$ & $0.006$ & $$-$0.004$ & $$-$0.005$ \\ 
			TE & $$-$0.005$ & $$-$\textbf{0.061}$ & $\textbf{0.057}$ & $$-$0.007$ & $0.016$ \\ 
			PE & $0$ & $$-$0.003$ & $0.006$ & $$-$0.002$ & $$-$0.001$ \\ 
			U & $$-$0.002$ & $$-$\textbf{0.025}$ & $0.005$ & $$-$0.012$ & $\textbf{0.034}$ \\ 
			IN & $$-$\textbf{0.007}$ & $$-$0.003$ & $0$ & $$-$\textbf{0.015}$ & $\textbf{0.024}$ \\ 
			\hline \\[-1.8ex] 
		\end{tabular} \\
	\end{subtable} 
	\caption*{\scriptsize{\textit{Note}: in bold we report probabilities, calculated via bootstrap (see Appendix \ref{app:bootstrapProcedures}), which are statistically significant at 5\% level.}}
\end{table}

The case of males looks very different. We observe a similar increase in the stock of permanent employment (+1.2 pp) and a reduction in temporary employment (-0.7 pp). However, we also report a reduction in unemployment (-0.9 pp) and an increase in the inactivity share (+0.9 pp) (Table \ref{tab:FittedForecastedNumbersMales}). These changes in the stocks are driven by increased flows from temporary to permanent employment (+5.7 pp), but also by increased flows from unemployment to inactivity (+3.4 pp), increased persistence into inactivity (+2.4 pp) and reduced outflows from inactivity (-1.5 pp) (Table \ref{tab:FittedForecastedTransProbMales}). Therefore, for the case of males, the increased rigidity of temporary contracts led to more stability, but at a cost of lower labour force participation.

\paragraph{Age}

 Among young workers (below the age of 35), the increase in the share of permanent employment (+2.2 pp) comes together with the decrease in the share of temporary employment (2.5 pp) (Table \ref{tab:FittedForecastedSharesYoung}). Looking at numbers (Table \ref{tab:FittedForecastedNumbersYoung}), this translates into 157.000 fewer temporary workers and 144,000 more permanent workers.
These trends are confirmed by increased transition probabilities from temporary to permanent employment (+9.1 pp)(Table \ref{tab:FittedForecastedTransProbYoung}). Among adults we observe  similar effects, but of smaller magnitude (Table \ref{tab:FittedForecastedSharesAdults} in Appendix \ref{app:heterogeneity}).
 
\begin{table}[!htbp]
	\scriptsize
		\caption{Fitted versus forecasted  statistics - Young.}
	\begin{subtable}[h]{\textwidth}
	\caption{Fitted versus forecasted  shares of individuals in different labour market states from quarter IV of 2018 to quarter IV of 2019- Young.}
	\label{tab:FittedForecastedSharesYoung}
	\centering
	\begin{tabular}{@{\extracolsep{5pt}} cccccc} 
		\\[-1.8ex]\hline 
		\hline \\[-1.8ex] 
		& SE & TE & PE & U & IN \\ 
		\hline \\[-1.8ex] 
Fitted  & $0.111$ & $\textbf{0.145}$ & $\textbf{0.386}$ & $0.088$ & $0.269$ \\
(s.e.) & ($0.003$) & ($0.004$) & ($0.004$) & ($0.003$) & ($0.005$) \\ 
Forecasted & $0.106$ & $\textbf{0.170}$ & $\textbf{0.364}$ & $0.091$ & $0.269$ \\ 
(s.e.) & ($0.003$) & ($0.005$) & ($0.005$) & ($0.006$) & ($0.006$) \\ 
		\hline 
	\end{tabular} 
\\
\end{subtable} \\
\vspace{0.5cm}

\begin{subtable}[htbp]{\textwidth}
	\caption{Difference between fitted and forecasted numbers of individuals in different labour market states from quarter IV of 2018 to quarter IV of 2019 - Young.}
	\label{tab:FittedForecastedNumbersYoung}
	\centering
	\begin{tabular}{@{\extracolsep{5pt}} cccccc} 
		\\[-1.8ex]\hline 
		\hline \\[-1.8ex] 
		& SE & TE & PE & U & IN \\ 
		\hline \\[-1.8ex] 
	C.I. 97.5\% & $73,104$ & $$-$73,369$ & $208,542$ & $63,173$ & $103,701$ \\ 
Difference & $32,434$ & $$-$\textbf{157,828}$ & $\textbf{144,349}$ & $$-$19,004$ & $49$ \\ 
C.I. 2.5\% & $$-$31,548$ & $$-$225,635$ & $46,658$ & $$-$88,141$ & $$-$88,173$ \\ 
		\hline \\[-1.8ex] 
	\end{tabular}
\vspace{0.3cm}
\end{subtable} \\

\begin{subtable}[htbp]{\textwidth}
	\caption{Fitted versus forecasted cumulative transition probabilities from quarter IV of 2018 to quarter IV of 2019 - Young.}
	\label{tab:FittedForecastedTransProbYoung}
	\centering 
	\begin{tabular}{@{\extracolsep{5pt}} cccccc} 
		\\[-1.8ex]\hline 
		\hline \\[-1.8ex] 
		& SE & TE & PE & U & IN \\ 
		\hline \\[-1.8ex] 
		SE & $0.040$ & $$-$0.011$ & $$-$0.002$ & $$-$0.004$ & $$-$0.023$ \\ 
TE & $0.001$ & $$-$\textbf{0.076}$ & $\textbf{0.091}$ & $$-$0.008$ & $$-$0.008$ \\ 
PE & $$-$0.001$ & $$-$0.005$ & $0.011$ & $$-$\textbf{0.006}$ & $0.001$ \\ 
U & $0.006$ & $$-$\textbf{0.032}$ & $0.013$ & $0.004$ & $0.009$ \\ 
IN & $0.002$ & $$-$\textbf{0.023}$ & $0.010$ & $0.002$ & $0.009$ \\
		\hline \\[-1.8ex] 
	\end{tabular} 
\end{subtable}
\caption*{\scriptsize{\textit{Note}: in bold we report probabilities, calculated via bootstrap (see Appendix \ref{app:bootstrapProcedures}), which are statistically significant at 5\% level.}}
\end{table}

\paragraph{Education level}

We then distinguish between workers with high and low education level. Among less educated, we observe an increase in permanent employment (+0.9 pp) and a decrease in the stock of temporary employment (-0.8 pp), corresponding to 263,000 fewer temporary workers and 286,000 more permanent workers (Tables \ref{tab:FittedForecastedSharesLE} and \ref{tab:FittedForecastedNumbersLE}). The same pattern is observed among high-educated workers but although the percentages are larger (+1.1 pp and -1.4 pp, respectively) the numbers are smaller (+75,000 and -91,000 respectively) (Tables \ref{tab:FittedForecastedSharesHE} and \ref{tab:FittedForecastedNumbersHE} in Appendix \ref{app:heterogeneity}). Among low-educated workers we also observe a decrease in unemployment, due to increased transitions from unemployment to inactivity (+2.1 pp) and reduced inflows from inactivity to unemployment (-0.9 pp).

\begin{table}[!htbp]
\caption{Fitted versus forecasted  statistics - Low-educated.}
\scriptsize
\begin{subtable}[h]{\textwidth}
\caption{Fitted versus forecasted  shares of individuals in different labour market states from quarter IV of 2018 to quarter IV of 2019- Low-educated.}
\label{tab:FittedForecastedSharesLE}
\centering
\begin{tabular}{@{\extracolsep{5pt}} cccccc} 
\\[-1.8ex]\hline 
\hline \\[-1.8ex] 
& SE & TE & PE & U & IN \\ 
\hline \\[-1.8ex] 
Fitted   & $0.171$ & $\textbf{0.089}$ & $\textbf{0.420}$ & $\textbf{0.059}$ & $0.260$ \\
(s.e.) & ($0.001$) & ($0.001$) & ($0.001$) & ($0.001$) & ($0.002$) \\ 
Forecasted  & $0.173$ & $\textbf{0.099}$ & $\textbf{0.411}$ & $\textbf{0.066}$ & $0.252$ \\   
(s.e.) & ($0.001$) & ($0.002$) & ($0.002$) & ($0.002$) & ($0.002$) \\ 
\hline 
\end{tabular} 
\\
\end{subtable} \\
\vspace{0.5cm}

\begin{subtable}[htbp]{\textwidth}
\caption{Difference between fitted and forecasted numbers of individuals in different labour market states from quarter IV of 2018 to quarter IV of 2019 - Low-educated.}
\label{tab:FittedForecastedNumbersLE}
\centering
\begin{tabular}{@{\extracolsep{5pt}} cccccc} 
\\[-1.8ex]\hline 
\hline \\[-1.8ex] 
& SE & TE & PE & U & IN \\ 
\hline \\[-1.8ex] 
		C.I. 97.5\% & $130,769$ & $$-$109,306$ & $436,588$ & $$-$42,996$ & $282,646$ \\ 
Difference & $37,230$ & $$-$\textbf{263,163}$ & $\textbf{286,377}$ & $$-$\textbf{169,367}$ & $108,922$ \\ 
C.I. 2.5\% & $$-$70,503$ & $$-$384,561$ & $129,230$ & $$-$290,958$ & $$-$92,901$ \\   
\hline \\[-1.8ex] 
\end{tabular}
\vspace{0.3cm} 
\end{subtable} \\
\begin{subtable}[htbp]{\textwidth}
\caption{Fitted versus forecasted cumulative transition probabilities from quarter IV of 2018 to quarter IV of 2019 - Low-educated.}
\label{tab:FittedForecastedTransProbLE}
\centering 
\begin{tabular}{@{\extracolsep{5pt}} cccccc} 
\\[-1.8ex]\hline 
\hline \\[-1.8ex] 
& SE & TE & PE & U & IN \\ 
\hline \\[-1.8ex] 
		SE & $0.008$ & $$-$\textbf{0.006}$ & $$-$0.001$ & $$-$0.001$ & $0$ \\ 
TE & $0.001$ & $$-$\textbf{0.053}$ & $\textbf{0.069}$ & $$-$\textbf{0.009}$ & $$-$0.007$ \\ 
PE & $0.002$ & $$-$0.003$ & $0.005$ & $$-$0.001$ & $$-$0.002$ \\ 
U & $$-$0.002$ & $$-$\textbf{0.017}$ & $0.005$ & $$-$0.007$ & $\textbf{0.021}$ \\ 
IN & $$-$0.001$ & $$-$0.003$ & $\textbf{0.004}$ & $$-$\textbf{0.009}$ & $0.009$ \\ 
\hline \\[-1.8ex] 
\end{tabular} 
\end{subtable}
\caption*{\scriptsize{\textit{Note}: in bold we report probabilities, calculated via bootstrap (see Appendix \ref{app:bootstrapProcedures}), which are statistically significant at 5\% level.}}
\end{table} 

\begin{table}[!htbp]
	\caption{Fitted versus forecasted  statistics - North.}
	\scriptsize
	\begin{subtable}[h]{\textwidth}
		\caption{Fitted versus forecasted  shares of individuals in different labour market states from quarter IV of 2018 to quarter IV of 2019- North.}
		\label{tab:FittedForecastedSharesNorth}
		\centering
		\begin{tabular}{@{\extracolsep{5pt}} cccccc} 
			\\[-1.8ex]\hline 
			\hline \\[-1.8ex] 
			& SE & TE & PE & U & IN \\ 
			\hline \\[-1.8ex] 
			Fitted  & $0.134$ & $\textbf{0.079}$ & $\textbf{0.443}$ & $\textbf{0.037}$ & $0.307$ \\
			(s.e.) & ($0.001$) & $(0.002$) & ($0.002$) & ($0.001$) & ($0.002$) \\ 
			Forecasted & $0.132$ & $\textbf{0.092}$ & $\textbf{0.429}$ & $\textbf{0.042}$ & $0.306$ \\ 
			(s.e.) & ($0.002$) & ($0.002$) & ($0.002$) & ($0.001$) & ($0.002$) \\  
			\hline 
		\end{tabular} 
		\\
		\end{subtable} \\
		\vspace{0.5cm}
		
		\begin{subtable}[h]{\textwidth}
			\caption{Difference between fitted and forecasted numbers of individuals in different labour market states from quarter IV of 2018 to quarter IV of 2019 - North.}
			\label{tab:FittedForecastedNumbersNorth}
			\centering
			\begin{tabular}{@{\extracolsep{5pt}} cccccc} 
				\\[-1.8ex]\hline 
				\hline \\[-1.8ex] 
				& SE & TE & PE & U & IN \\ 
				\hline \\[-1.8ex] 
				C.I. 97.5\% & $133,490$ & $$-$179,973$ & $465,266$ & $$-$38,978$ & $201,467$ \\ 
				Difference & $46,377$ & $$-$\textbf{316,343}$ & $\textbf{355,082}$ & $$-$\textbf{127,210}$ & $42,095$ \\ 
				C.I. 2.5\% & $$-$59,463$ & $$-$437,264$ & $212,886$ & $$-$197,879$ & $$-$95,701$ \\   
				\hline \\[-1.8ex] 
			\end{tabular}
			\vspace{0.3cm}
		\end{subtable} \\
		
		\begin{subtable}[h]{\textwidth}
			\caption{Fitted versus forecasted cumulative transition probabilities from quarter IV of 2018 to quarter IV of 2019 - North.}
			\label{tab:FittedForecastedTransProbNorth}
			\centering 
			\begin{tabular}{@{\extracolsep{5pt}} cccccc} 
				\\[-1.8ex]\hline 
				\hline \\[-1.8ex] 
				& SE & TE & PE & U & IN \\ 
				\hline \\[-1.8ex] 
				SE & $0.002$ & $$-$0.005$ & $0.007$ & $0$ & $$-$0.004$ \\ 
				TE & $0.001$ & $$-$\textbf{0.089}$ & $\textbf{0.110}$ & $$-$0.005$ & $$-$0.017$ \\ 
				PE & $0.002$ & $$-$\textbf{0.005}$ & $0.004$ & $$-$0.001$ & $$-$0.001$ \\ 
				U & $$-$0.001$ & $$-$\textbf{0.026}$ & $\textbf{0.027}$ & $$-$0.007$ & $0.007$ \\ 
				IN & $0.002$ & $$-$0.005$ & $0.004$ & $$-$\textbf{0.013}$ & $\textbf{0.013}$ \\  
				\hline \\[-1.8ex] 
			\end{tabular} 
		\end{subtable}
		\caption*{\scriptsize{\textit{Note}: in bold we report probabilities, calculated via bootstrap (see Appendix \ref{app:bootstrapProcedures}), which are statistically significant at 5\% level.}}
	\end{table}

\paragraph{Geography}

Finally, we distinguish between workers living in the North and Center versus the South of the country. While in the North we observe a drop in temporary employment (-1.3 pp) and unemployment (-0.5 pp) and an increase in permanent employment (+1.4 pp) (Table \ref{tab:FittedForecastedSharesNorth}), we do not find any effect in the South (Table \ref{tab:FittedForecastedSharesSouth} in Appendix \ref{app:heterogeneity}). In the North, the lower number of workers in unemployment is due to a decreased inflow of workers from the inactivity state (-1.3 pp) and an increase outflow of workers from unemployment to permanent employment (+2.7 pp). However, we also report higher persistence in inactivity (+1.3 pp), suggesting negative effects on labour force participation (Table \ref{tab:FittedForecastedTransProbNorth}).

\section{Concluding remarks \label{sec:concludingRemarks}}

In this paper, we evaluate the effects of a labour market reform implemented in Italy to increase the rigidity of temporary contracts (\textit{Decreto Dignit\`a}). 
We find a strong increase in the share of permanent employment, associated with a strong decrease in the share of temporary employment. Specifically, we find strong flows of workers moving from temporary to permanent employment, which was set as the main goal of the reform, which compensate the reduced persistence in temporary employment. We find the effect to be particularly strong among women, and young workers living in the North of Italy. In addition, we find the stock of individuals in unemployment to be significantly lower after the reform, likely due to a composition effect. Among specific categories of workers, such as males and low-educated workers, we also find some evidence of  reduced transitions from inactivity towards the labour market and increased persistence in inactivity.

These findings suggest that for the case of Italy the short-term benefits of the roll back in terms of employment and unemployment have been large, while the associated costs in terms of reduced inflows towards the labour market seem to have been relatively low.  It is important to remark that the \textit{Decreto Dignit\`a} was implemented after the introduction of the \textit{Jobs Act}, which introduced for permanent contracts a system of firing costs increasing with tenure. This latter reform significantly reduced the employment protection gap between the two types of contracts, increasing their substitutability. The roll back reform might have therefore been particularly successful due to its timing, as the cost for firms to swap from temporary to permanent contracts had already been reduced. One caveat of our study has to do with the observations being limited to the short-term (one year after the implementation of the reform). Since it would be difficult to isolate the effect of the reform from other confounding effects should have we decided to extend the period of analysis, we cannot foresee what will happen in the medium/longer run. 

Our findings  suggest that the firms' utilization of temporary contracts is often not justified by the temporary nature of the tasks. Therefore, other factors such as the  bargaining  power of firms and the consequent division of surplus when hiring workers on temporary versus permanent basis might play a significant role in the firms' choice of contracts \citep{card2022set}. 

 These results might also apply to other dual economies, such as Spain, France and Portugal, which have recently undertaken such roll back process to reduce labour segmentation \citep{cahuc2022employment, bentolila2019dual}.
However, our evidence seems to go against the one from Spain, where a number of countervailing reforms which have been implemented since the late 90s to reduce the utilization of temporary contracts have been only partially successful \citep{felgueroso2018surge, bentolila2008two, bentolila2012reforming}. 
More research to inform policy makers on the reversibility of roll back policies in dual economies is therefore paramount at this point in time.

\clearpage

\bibliographystyle{chicago}
\bibliography{references}

\newpage

\clearpage

\appendix

\begin{Large}
	\textbf{Appendix}
\end{Large}

\section{Bootstrap procedure for labour shares and transition probabilities \label{app:bootstrapProcedures}}

Given a sample of transitions $X$ of cardinality $N$ with sample weights $w^X$, the bootstrap procedure is composed of three steps \citep[Chapter 6]{efron1994introduction}:
\begin{enumerate}
\item Draw $B$ samples of cardinality $N$ by sampling with replacement from $X$ with sample weights $w^X$;
\item For every bootstrapped sample $b$ estimate matrix $\mathbf{M}^b$;
\item Compute the standard errors of the transition probabilities ${m}\left(i,j\right)$, $\sigma_{m\left(i,j\right)}$ as:
\[
\sigma_{m\left(i,j\right)} = \sqrt{\sum_{b=1}^{B} \dfrac{ \left[ {m}\left(i,j\right)^b - \overline{{m}}\left(i,j\right)^b\right]^2}{B}},
\]
where $ {m}\left(i,j\right)^b$ is the $(i,j)$ element of ${\mathbf{M}}^b$ and $\overline{{m}}\left(i,j\right)^b$ is the average $(i,j)$ element of all the $B$ bootstraps.
\end{enumerate}

The standard errors of shares are calculated in the same fashion, by sampling with replacement with sample weights from the original dataset of labour market states in each quarter.
The test of zero difference between two transition probabilities and/or between labour market shares is based on the bootstrap procedure suggested in \citet[Chapter 16]{efron1994introduction}.

\clearpage

\subsection{Robustness}\label{app:robustness}

\subsubsection{Placebo estimates}\label{app:placebo}

As a robustness exercise, we perform a placebo test where we use data from quarter I of 2016 to quarter III of 2017 to build the forecast for the following three quarters (from quarter IV of 2017 to quarter II of 2018). We then compare the fitted with the forecasted values. If our methodology is robust, we should observe no statistical difference between the two.

\begin{table}[!htbp] \centering 
\scriptsize
\caption{Placebo test.  Quarters used in the forecast: 2016.I - 2017.III  and forecasted quarters: 2017.IV - 2018.II.} 
\label{placebo}
\begin{subtable}[h]{\textwidth}
\centering
\caption{Fitted versus forecasted  shares of individuals in different labour market states from quarter IV of 2017 to quarter II of 2018.} 

\begin{tabular}{@{\extracolsep{5pt}} cccccc} 
\\[-1.8ex]\hline 
\hline \\[-1.8ex] 
& SE & TE & PE & U & IN \\ 
\hline \\[-1.8ex] 
Fitted  & $0.124$ & $0.084$ & $0.377$ & $0.066$ & $0.349$ \\
(s.e.) & $(0.001)$ & $(0.001)$ & $(0.001)$ & $(0.001)$ & $(0.002)$ \\ 
Forecasted & $0.123$ & $0.084$ & $0.373$ & $0.069$ & $0.352$ \\ 
(s.e.) & $(0.001)$ & $(0.002)$ & $(0.002)$ & $(0.001)$ & $(0.002)$ \\ 
\hline \\[-1.8ex] 
\end{tabular} 
\vspace{0.4cm}
\end{subtable}
\begin{subtable}[h]{\textwidth}
\centering
\caption{Difference between fitted and forecasted numbers of individuals in different labour market states from quarter IV of 2017 to quarter II of 2018.} 
\begin{tabular}{@{\extracolsep{5pt}} cccccc} 
\\[-1.8ex]\hline 
\hline \\[-1.8ex] 
& SE & TE & PE & U & IN \\ 
\hline \\[-1.8ex] 
C.I. 97.5\% & $155,872$ & $166,490$ & $291,794$ & $10,088$ & $56,365$ \\ 
Difference & $47,624$ & $20,911$ & $167,225$ & $$-$112,314$ & $$-$123,446$ \\ 
C.I. 2.5\% & $$-$48,742$ & $$-$102,950$ & $$-$9,862$ & $$-$220,637$ & $$-$302,577$ \\  
\hline 
\end{tabular} 
\end{subtable}
\caption*{\scriptsize{\textit{Note}: in bold we report probabilities, calculated via bootstrap (see Appendix \ref{app:bootstrapProcedures}), which are statistically significant at 5\% level.}}
\end{table} 

In Table \ref{placebo} we report the fitted and forecasted shares of individuals in different labour market states (panel a) and the difference between the fitted and forecasted numbers of individuals in different labour market shares (panel b). The presence of no statistically significant values supports the validity of our approach. 

\clearpage

\subsubsection{Quarter III of 2018}\label{app:Q32018}

The \textit{Decreto Dignit\`a} was implemented in the beginning of August 2018, i.e., in the middle of quarter III of 2018. There is therefore an issue of whether to include the quarter in the period of observation or in the forecasted period. Moreover, we observe quite large, but not persistent changes in the unemployment and inactivity shares in that quarter, which  slightly affect the results. While we have chosen to include the quarter in the period of observation in our preferred estimation, we report here a robustness exercise where instead quarter III of 2018 is included in the forecasted period.

\begin{table}[!htbp] \centering 
\scriptsize
	\caption{Counterfactual labour shares.  Quarters used in the forecast: 2016.I - 2018.II  and forecasted quarters: 2018.III - 2019.II.} 
 \begin{subtable}[h]{\textwidth}
 \centering
	\caption{Fitted versus forecasted  shares of individuals in different labour market states from quarter III of 2018 to quarter II of 2019.} 
	\begin{tabular}{@{\extracolsep{5pt}} cccccc} 
		\\[-1.8ex]\hline 
		\hline \\[-1.8ex] 
		& SE & TE & PE & U & IN \\ 
		\hline \\[-1.8ex] 
		Fitted  & $0.124$ & $\textbf{0.080}$ & $\textbf{0.382}$ & $\textbf{0.060}$ & $\textbf{0.355}$ \\
		(s.e.) & $(0.001)$ & $(0.001)$ & $(0.001)$ & $(0.001)$ & $(0.002)$ \\ 
		Forecasted & $0.123$ & $\textbf{0.092}$ & $\textbf{0.375}$ & $\textbf{0.065}$ & $\textbf{0.345}$ \\  
		(s.e.) & $(0.001)$ & $(0.002)$ & $(0.002)$ & $(0.001)$ & $(0.002)$ \\ 
		\hline \\[-1.8ex] 
		Change& 0.001&-\textbf{0.010}&+\textbf{0.006}&-\textbf{0.005}&+\textbf{0.008}\\
		\hline 
	\end{tabular} 
	 \vspace{0.4cm}
	\end{subtable}

 \begin{subtable}[h]{\textwidth}
 \centering
	\caption{Difference between fitted and forecasted numbers of individuals in different labour market states from quarter III of 2018 to quarter II of 2019.} 
	\begin{tabular}{@{\extracolsep{5pt}} cccccc} 
		\\[-1.8ex]\hline 
		\hline \\[-1.8ex] 
		& SE & TE & PE & U & IN \\ 
		\hline \\[-1.8ex] 
		C.I. 97.5\% & $174,200$ & $$-$246,421$ & $425,991$ & $$-$107,648$ & $558,412$ \\ 
Difference & $53,765$ & $$-$\textbf{459,261}$ & $\textbf{255,416}$ & $$-$\textbf{211,836}$ & $\textbf{361,916}$ \\ 
C.I. 2.5\% & $$-$56,056$ & $$-$595,701$ & $24,097$ & $$-$339,029$ & $149,208$ \\ 
		\hline \\[-1.8ex] 
	\end{tabular} 
\vspace{0.4cm}
\end{subtable}

\begin{subtable}[h]{\textwidth}
	\centering
	\caption{Fitted versus forecasted cumulative transition probabilities from quarter III of 2018 to quarter II of 2019.} 
	\begin{tabular}{@{\extracolsep{5pt}} cccccc} 
		\\[-1.8ex]\hline 
		\hline \\[-1.8ex] 
		& SE & TE & PE & U & IN \\ 
		\hline \\[-1.8ex] 
		SE & $\textbf{0.020}$ & $$-$\textbf{0.007}$ & $0$ & $$-$\textbf{0.006}$ & $$-$0.007$ \\ 
TE & $$-$0.001$ & $$-$\textbf{0.055}$ & $\textbf{0.051}$ & $$-$\textbf{0.008}$ & $0.013$ \\ 
PE & $0.001$ & $$-$0.003$ & $0.002$ & $$-$0.001$ & $0.001$ \\ 
U & $$-$0.006$ & $$-$\textbf{0.022}$ & $\textbf{0.018}$ & $$-$\textbf{0.013}$ & $\textbf{0.022}$ \\ 
IN & $$-$0.003$ & $$-$\textbf{0.012}$ & $0.001$ & $$-$\textbf{0.008}$ & $\textbf{0.021}$ \\  
		\hline \\[-1.8ex] 
	\end{tabular} 
	\end{subtable}
	\caption*{\scriptsize{\textit{Note}: in bold we report probabilities, calculated via bootstrap (see Appendix \ref{app:bootstrapProcedures}), which are statistically significant at 5\% level.}}
\end{table} 

\clearpage

\section{Heterogeneity}\label{app:heterogeneity}

\subsection{Adults}

\begin{table}[!htbp]
\scriptsize
\caption{Fitted versus forecasted  statistics - Adults.}
\begin{subtable}[h]{\textwidth}
\caption{Fitted versus forecasted  shares of individuals in different labour market states from quarter IV of 2018 to quarter IV of 2019- Adults.}
\label{tab:FittedForecastedSharesAdults}
\centering
\begin{tabular}{@{\extracolsep{5pt}} cccccc} 
\\[-1.8ex]\hline 
\hline \\[-1.8ex] 
& SE & TE & PE & U & IN \\ 
\hline \\[-1.8ex] 
Fitted & $0.151$ & $\textbf{0.054}$ & $\textbf{0.446}$ & $0.046$ & $0.302$ \\  
(s.e) & ($0.001$) & ($0.001$) & ($0.002$) & ($0.001$) & ($0.002$) \\ 
Forecasted & $0.152$ & $\textbf{0.058}$ & $\textbf{0.442}$ & $0.048$ & $0.301$ \\  
(s.e) & ($0.001$) & ($0.001$) & ($0.002$) & ($0.001$) & ($0.002$) \\  
\hline 
\end{tabular} 
\\
\end{subtable} \\
\vspace{0.4cm}

\begin{subtable}[h]{\textwidth}
\caption{Difference between fitted and forecasted numbers of individuals in different labour market states from quarter IV of 2018 to quarter IV of 2019 - Adults.}
\label{tab:FittedForecastedNumbersAdults}
\centering
\begin{tabular}{@{\extracolsep{5pt}} cccccc} 
\\[-1.8ex]\hline 
\hline \\[-1.8ex] 
& SE & TE & PE & U & IN \\ 
\hline \\[-1.8ex] 
C.I. 97.5\% & $68,262$ & $$-$3,409$ & $252,026$ & $42,681$ & $177,694$ \\ 
Difference & $$-$18,344$ & $$-$\textbf{101,740}$ & $\textbf{126,790}$ & $$-$27,858$ & $21,152$ \\ 
C.I. 2.5\% & $$-$106,247$ & $$-$205,855$ & $$-$5,634$ & $$-$123,447$ & $$-$96,556$ \\  
\hline \\[-1.8ex] 
\end{tabular}
\vspace{0.4cm} 
\end{subtable} \\

\begin{subtable}[h]{\textwidth}
\caption{Fitted versus forecasted cumulative transition probabilities from quarter IV of 2018 to quarter IV of 2019 - Adults.}
\label{tab:FittedForecastedTransProbAdults}
\centering 
\begin{tabular}{@{\extracolsep{5pt}} cccccc} 
\\[-1.8ex]\hline 
\hline \\[-1.8ex] 
& SE & TE & PE & U & IN \\ 
\hline \\[-1.8ex] 
SE & $0.003$ & $$-$0.002$ & $0$ & $$-$0.001$ & $0$ \\ 
TE & $$-$0.004$ & $$-$\textbf{0.049}$ & $\textbf{0.063}$ & $0$ & $$-$0.010$ \\ 
PE & $0.001$ & $$-$0.001$ & $0$ & $0$ & $0$ \\ 
U & $$-$0.009$ & $$-$0.008$ & $\textbf{0.015}$ & $$-$0.003$ & $0.005$ \\ 
IN & $$-$0.004$ & $$-$0.001$ & $0.003$ & $$-$0.003$ & $0.005$ \\  
\hline \\[-1.8ex] 
\end{tabular} 
\end{subtable}
\caption*{\scriptsize{\textit{Note}: in bold we report probabilities, calculated via bootstrap (see Appendix \ref{app:bootstrapProcedures}), which are statistically significant at 5\% level.}}
\end{table}

\clearpage

\subsection{High-educated}

\begin{table}[!htbp]
\scriptsize
\caption{Fitted versus forecasted  statistics - High-educated.}
\begin{subtable}[h]{\textwidth}
\caption{Fitted versus forecasted  shares of individuals in different labour market states from quarter IV of 2018 to quarter IV of 2019- High-educated.}
\label{tab:FittedForecastedSharesHE}
\centering
\begin{tabular}{@{\extracolsep{5pt}} cccccc} 
\\[-1.8ex]\hline 
\hline \\[-1.8ex] 
& SE & TE & PE & U & IN \\ 
\hline \\[-1.8ex] 
Fitted & $0.197$ & $\textbf{0.076}$ & $\textbf{0.524}$ & $0.040$ & $0.163$ \\
(s.e) & ($0.003$) & ($0.003$) & ($0.003$) & ($0.002$) & ($0.004$) \\ 
Forecasted & $0.198$ & $\textbf{0.090}$ & $\textbf{0.513}$ & $0.039$ & $0.161$ \\  
(s.e) & ($0.003$) & ($0.003$) & ($0.004$) & ($0.003$) & ($0.004$) \\ 
\hline 
\end{tabular} 
\\

\end{subtable} \\
\vspace{0.5cm}

\begin{subtable}[htbp]{\textwidth}
\caption{Difference between fitted and forecasted numbers of individuals in different labour market states from quarter IV of 2018 to quarter IV of 2019 - High-educated.}
\label{tab:FittedForecastedNumbersHE}
\centering
\begin{tabular}{@{\extracolsep{5pt}} cccccc} 
\\[-1.8ex]\hline 
\hline \\[-1.8ex] 
& SE & TE & PE & U & IN \\ 
\hline \\[-1.8ex] 
C.I. 97.5\% & $47,255$ & $$-$34,760$ & $136,163$ & $50,423$ & $92,354$ \\ 
Difference & $$-$5,518$ & $$-$\textbf{91,898}$ & $\textbf{75,765}$ & $10,266$ & $11,385$ \\ 
C.I. 2.5\% & $$-$57,957$ & $$-$151,868$ & $6,300$ & $$-$36,759$ & $$-$48,263$ \\
\hline \\[-1.8ex] 
\end{tabular}
\vspace{0.3cm}
\end{subtable} \\

\begin{subtable}[htbp]{\textwidth}
\caption{Fitted versus forecasted cumulative transition probabilities from quarter IV of 2018 to quarter IV of 2019 - High-educated.}
\label{tab:FittedForecastedTransProbYoung_II}
\centering 
\begin{tabular}{@{\extracolsep{5pt}} cccccc} 
\\[-1.8ex]\hline 
\hline \\[-1.8ex] 
& SE & TE & PE & U & IN \\ 
\hline \\[-1.8ex] 
SE & $0.016$ & $$-$0.002$ & $$-$0.001$ & $$-$0.003$ & $$-$0.009$ \\ 
TE & $$-$0.008$ & $$-$\textbf{0.117}$ & $\textbf{0.103}$ & $0.019$ & $0.003$ \\ 
PE & $$-$0.003$ & $$-$0.003$ & $0.005$ & $$-$\textbf{0.003}$ & $0.003$ \\ 
U & $$-$0.014$ & $$-$0.024$ & $\textbf{0.036}$ & $0.005$ & $$-$0.004$ \\ 
IN & $$-$0.006$ & $$-$0.010$ & $$-$0.004$ & $0.010$ & $0.010$ \\     
\hline \\[-1.8ex] 
\end{tabular} 
\end{subtable}
\caption*{\scriptsize{\textit{Note}: in bold we report probabilities, calculated via bootstrap (see Appendix \ref{app:bootstrapProcedures}), which are statistically significant at 5\% level.}}
\end{table} 

\clearpage

\subsection{South}

\begin{table}[!htbp]
\scriptsize
\caption{Fitted versus forecasted  statistics - South.}
\begin{subtable}[h]{\textwidth}
\caption{Fitted versus forecasted  shares of individuals in different labour market states from quarter IV of 2018 to quarter IV of 2019- South.}
\label{tab:FittedForecastedSharesSouth}
\centering
\begin{tabular}{@{\extracolsep{5pt}} cccccc} 
\\[-1.8ex]\hline 
\hline \\[-1.8ex] 
& SE & TE & PE & U & IN \\ 
\hline \\[-1.8ex] 
Fitted   & $0.108$ & $0.082$ & $0.261$ & $0.086$ & $0.463$ \\ 
(s.e) & ($0.002$) & ($0.002$) & ($0.002$) & ($0.002$) & ($0.003$) \\ 
Forecasted & $0.108$ & $0.085$ & $0.257$ & $0.090$ & $0.460$ \\  
(s.e) & ($0.002$) & ($0.002$) & ($0.002$) & ($0.004$) & ($0.005$) \\ 
\hline 
\end{tabular} 
\\
\vspace{0.4cm} 
\end{subtable} \\

\begin{subtable}[h]{\textwidth}
\caption{Difference between fitted and forecasted numbers of individuals in different labour market states from quarter IV of 2018 to quarter IV of 2019 - South.}
\label{tab:FittedForecastedNumbersSouth}
\centering
\begin{tabular}{@{\extracolsep{5pt}} cccccc} 
\\[-1.8ex]\hline 
\hline \\[-1.8ex] 
& SE & TE & PE & U & IN \\ 
\hline \\[-1.8ex] 
C.I. 97.5\% & $51,081$ & $49,026$ & $137,716$ & $84,234$ & $170,891$ \\ 
Difference & $$-$4,147$ & $$-$39,886$ & $54,332$ & $$-$50,372$ & $40,074$ \\ 
C.I. 2.5\% & $$-$76,266$ & $$-$129,303$ & $$-$33,073$ & $$-$141,394$ & $$-$122,738$ \\ 
\hline \\[-1.8ex] 
\end{tabular}
\vspace{0.4cm} 
\end{subtable} \\
\begin{subtable}[h]{\textwidth}
\caption{Fitted versus forecasted cumulative transition probabilities from quarter IV of 2018 to quarter IV of 2019 - South.}
\label{tab:FittedForecastedTransProbYoung_III}
\centering 
\begin{tabular}{@{\extracolsep{5pt}} cccccc} 
\\[-1.8ex]\hline 
\hline \\[-1.8ex] 
& SE & TE & PE & U & IN \\ 
\hline \\[-1.8ex] 
SE & $\textbf{0.032}$ & $$-$0.006$ & $$-$0.013$ & $$-$0.007$ & $$-$0.006$ \\ 
TE & $$-$0.004$ & $$-$0.016$ & $0.024$ & $$-$0.006$ & $0.002$ \\ 
PE & $$-$0.003$ & $0.001$ & $0.006$ & $$-$0.003$ & $$-$0.002$ \\ 
U & $$-$0.006$ & $$-$0.007$ & $0$ & $$-$0.003$ & $0.015$ \\ 
IN & $$-$\textbf{0.005}$ & $$-$0.002$ & $0.004$ & $$-$0.004$ & $0.006$ \\   
\hline \\[-1.8ex] 
\end{tabular} 
\end{subtable}
\caption*{\scriptsize{\textit{Note}: in bold we report probabilities, calculated via bootstrap (see Appendix \ref{app:bootstrapProcedures}), which are statistically significant at 5\% level.}}
\end{table} 

\clearpage

\section{Composition effect in the transitions between labour market states driven by a Markovian process \label{app:transitionBetweenStatus}}

In this section we demonstrate that the stock of unemployed individuals might decrease even  when inflow and outflow rates to and from unemployment are not affected by a policy, i.e. the  \textit{Decreto Dignit\`a}, due to a composition effect.

Consider an economy with a constant mass $\bar{L}$ of individuals who can be in any of the three labour market states: temporary contract ($T$), permanent contract ($P$) or unemployed ($U$), such that:
\begin{equation}\label{LF}
	T_t+P_t+U_t = \bar{L}.
\end{equation}
Assume that the individuals can freely move across the states at some transition rates which are constant over time, i.e. the labour market dynamics can be described by a Markovian process. The observed dynamics are the outcome of inflows and outflows of individuals across the three labour market states and are fully described by the set of transitions rates, i.e.,:
\begin{equation} \label{eq:MarvovianLaborMarketDynamics}
	\begin{bmatrix} 
		\pi(T)_{t+1} \\
		\pi(P)_{t+1} \\
		\pi(U)_{t+1} 
	\end{bmatrix} =
\begin{bmatrix} 
	m(T,T) & m(P,T) & m(U,T) \\
	m(T,P) & m(P,P) & m(U,P) \\
	m(T,U) & m(P,U) & m(U,U)
\end{bmatrix}
\times
	\begin{bmatrix} 
	\pi(T)_{t}\\
	\pi(P)_{t} \\
	\pi(U)_{t}
	\end{bmatrix},
\end{equation}	
where $\pi(T)_{t}\equiv T_t/\bar{L}$, $\pi(P)_{t} \equiv P_t/\bar{L}$, and $\pi(U)_{t} \equiv U_t/\bar{L}$ are  the share of individuals in the T, P, and U states, respectively ( $\pi(T)_{t}+ \pi(P)_{t} + \pi(U)_{t} =1$), while $m(i,j) \geq 0$ is the probability to transit from state $i$ to state $j$ in the period $[t,t+1]$ ($m(i,T)_{t} + m(i,P)_{t} + m(i,U)_{t} =1$ for $i \in \left\{T,P,U\right\}$).
Under general conditions (Markov chain is irreducible and aperiodic), there exist an equilibrium distribution of individuals across the three states, independent of the initial distribution. The equilibrium share of individuals in unemployment is given by:

\begin{eqnarray}
		\label{equ} 
	 \pi_U^{EQ}=\dfrac{m(T,U)(1-m(P,P))+m(T,P)m(P,U)  }{\splitfrac{\textstyle (1-m(P,P))(1+m(T,U)-m(U,U))+m(T,P)(1+m(P,U)-m(U,U))}{\textstyle-m(U,P)(m(P,U)}}. 
\end{eqnarray}
 
Taking the derivative of Equation \ref{equ} with respect to the probability of transiting from temporary to permanent employment, $m(T,P)$, which is the probability directly affected by the \textit{Decreto Dignit\`a}, we get:

\begin{equation}\label{partialPTP}
\frac{\partial \pi_U^{EQ}}{\partial m(T,P)}=\dfrac{\left(m(P,U)-m(T,U)\right)\left[(1-m(P,P))m(U,T)+m(U,P)m(P,T)\right]} {\splitfrac{\textstyle \left[(1-m(P,P))(1+m(T,U)-m(U,U))+m(T,P)(1+m(P,U)-m(U,U))\right.}{\textstyle \left.-m(U,P)(m(P,U)-m(T,U))\right]^2}}.
\end{equation}

From Equation \ref{partialPTP}, we observe that the sign of the derivative depends only on the sign of the term $m(P,U)-m(T,U)$, which is expected to be negative as the probability to transit to unemployment from permanent employment is smaller than the probability to transit to unemployment from temporary employment. Therefore, the share of individuals in unemployment may actually decrease simply due to a larger share of individuals moving from temporary to permanent employment, i.e., a composition effect, with the probabilities to enter and exit the unemployment pool being unchanged, as it was the case of the \textit{Decreto Dignit\`a}.

\end{document}